\documentstyle[preprint,eqsecnum,aps,prd,tighten]{revtex}

\input boxedeps
\SetepsfEPSFSpecial
\HideDisplacementBoxes
\newcommand{\etal}{{\em et al.}}

\newcommand{\cf}{{\em cf.\ }}
\newcommand{\gevcc}{\hbox{ GeV}\!/\!c^2}
\newcommand{\gev}{\hbox{ GeV}}
\newcommand{\ev}{\hbox{ eV}}

\newcommand{\tev}{\hbox{ TeV}}
\newcommand{\pev}{\hbox{ PeV}}
\newcommand{\cm}{\hbox{ cm}}
\newcommand{\km}{\hbox{ km}}
\newcommand{\cmwe}{\hbox{ cmwe}}
\newcommand{\kmwe}{\hbox{ kmwe}}
\newcommand{\Flux}{\hbox{ km}^{-2}\hbox{ s}^{-1}\hbox{ sr}^{-1}}
\newcommand{\flux}{\hbox{ cm}^{-2}\hbox{ s}^{-1}\hbox{ sr}^{-1}\gev^{-1}}
\newcommand{\eqn}[1]{(\ref{#1})}
	\newcommand{\Qpl}[3]{{Phys. Lett.} {\bf #1,} #2 (19#3)}						 %
	\newcommand{\Qprl}[3]{Phys. Rev. Lett. {\bf #1,} #2 (19#3)}				 %
	\newcommand{\prep}[3]{Phys. Rep. {\bf #1,} #2 (19#3)}					 %
	\newcommand{\rpp}[3]{Rep. Prog. Phys. {\bf #1,} #2 (19#3)}				 %
	\newcommand{\Qpr}[3]{{Phys. Rev. D}{\bf #1,} #2 (19#3)}						 %
	\newcommand{\np}[3]{Nucl. Phys. {\bf #1,} #2 (19#3)}						 %
	\newcommand{\npbps}[3]{Nucl. Phys.	B (Proc. Supp.) {\bf #1,} #2 (19#3)}	 %
	\newcommand{\zp}[3]{Z.~Phys. C{\bf#1,} #2 (19#3)}						 %
	\newcommand{\astropp}[3]{Astropart. Phys. {\bf #1,} #2 (19#3)}			 %
	\newcommand{\ib}[3]{{\em ibid.\/} {\bf #1,} #2 (19#3)}							 %
	\newcommand{\Qnat}[3]{Nature (London) {\bf #1,} #2 (19#3)}				 %
	\newcommand{\nuovocim}[3]{Nuovo Cim. {\bf #1,} #2 (19#3)}				 %
	\newcommand{\yadfiz}[4]{Yad. Fiz. {\bf #1,} #2 (19#3) [English			 %
	transl.: Sov. J. Nucl.	Phys. {\bf #1,} #4 (19#3)]}						 %
	\newcommand{\philt}[3]{Phil. Trans. Roy. Soc. London A {\bf	#1,} #2  
	(19#3)}																		 %
	\newcommand{\hepph}[1]{(electronic archive:	hep--ph/#1)}						 %
	\newcommand{\hepex}[1]{(electronic archive:	hep--ex/#1)}						 %
	\newcommand{\astro}[1]{(electronic archive:	astro--ph/#1)}						 %
\begin{document}
%\draft
\preprint{FERMILAB--PUB--98/087--T}
\title{Neutrino Interactions at Ultrahigh Energies}
\author{Raj Gandhi\thanks{E-mail: \textsf{raj@mri.ernet.in}}}
\address{Mehta Research Institute \\ Chhatnag Road, Jhusi, Allahabad 
211019, India}
\author{Chris Quigg\thanks{E-mail: \textsf{quigg@fnal.gov}}}
\address{Theoretical Physics Department, 
Fermi National Accelerator 
Laboratory \\ P.O.\ Box 500, Batavia, Illinois 60510 USA}
\author{Mary Hall Reno\thanks{E-mail: 
\textsf{reno@hepsun1.physics.uiowa.edu}}}
\address{Department of Physics and Astronomy, University of Iowa \\ Iowa 
City, Iowa 52242 USA}
\author{Ina Sarcevic\thanks{E-mail: 
\textsf{ina@gluon.physics.arizona.edu}}}
\address{Department of Physics, University of Arizona \\ Tucson, Arizona 
85721 USA}
\date{\today}
\maketitle
\begin{abstract}
    We report new calculations of the cross sections for deeply inelastic 
    neutrino-nucleon scattering at neutrino energies between $10^{9}\ev$ 
    and $10^{21}\ev$.  We compare with results in the literature and 
    assess the reliability of our predictions.  For completeness, we 
    briefly review the cross sections for neutrino interactions with 
    atomic electrons, emphasizing the role of the $W$-boson resonance in 
    $\bar{\nu}_{e}e$ interactions for neutrino energies in the 
    neighborhood of $6.3\pev$.  Adopting model predictions for 
    extraterrestrial neutrino fluxes from active galactic nuclei, 
    gamma-ray bursters, and the collapse of topological defects, we 
    estimate event rates in large-volume water \v{C}erenkov detectors and 
    large-area ground arrays.
\end{abstract}

\narrowtext

\section{Introduction}\label{sec:intro}
Neutrino observatories hold great promise for probing the deepest 
reaches of stars and galaxies \cite{Totsuka,GHS,BahWolf,bhp}.  Unlike 
charged particles, neutrinos arrive on a direct line from their 
source, undeflected by magnetic fields.  Unlike photons, neutrinos 
interact weakly, so they can penetrate thick columns of matter.
For example, the 
interaction length of a 1-TeV neutrino is about 2.5 million kilometers 
of water, or 250 kilotonnes/cm$^{2}$, whereas high-energy photons are 
blocked by a few hundred grams/cm$^{2}$.

Ultrahigh-energy neutrinos can be detected by observing long-range 
muons produced in charged-current neutrino-nucleon interactions.  To 
reduce the background from muons produced in the atmosphere, it is 
advantageous to site a neutrino telescope at a depth of several 
kilometers (water equivalent) or to observe upward-going muons.  The 
reactions $(\nu_{\ell},\bar{\nu}_{\ell}) N \rightarrow 
(\ell^{-},\ell^{+}) + \hbox{anything}$ and 
$(\nu_{\ell},\bar{\nu}_{\ell}) N \rightarrow 
(\nu_{\ell},\bar{\nu}_{\ell}) + \hbox{anything}$ are the major sources 
of both the desired signal and the attenuation of the neutrino 
``beam'' as it passes through the Earth \textit{en route} to the 
detector.  Through the past dozen years, improving knowledge of the 
partonic structure of the nucleon has made possible a series of 
increasingly refined predictions for the interaction cross sections 
\cite{McKR,QRW,RQ,BKKLZ,FMcKR,pz,Zhel,GQRS,ghill}.  Over the same 
period, ideas about the flux of neutrinos from active galactic nuclei 
(AGNs) and other extraterrestrial sources have evolved considerably.  
The observation \cite{1987A} of neutrinos correlated with supernova 
SN1987A and the detection of solar neutrinos by observing the 
direction of recoil electrons from neutrino interactions 
\cite{Kamio,SuperKsoleil} showed the promise of neutrino observatories 
for astrophysical studies.  The detection of neutrinos produced by 
cosmic-ray interactions in the Earth's atmosphere 
\cite{IMBAtm,KamioAtm,Frejus,frelim} has emerged as a tool for 
investigating neutrino oscillations 
\cite{soudan2,kearns,SuperKanom,SuperKatm}.  Plans for neutrino 
observatories that will detect neutrinos that originate beyond Earth 
have matured to the point that it is now reasonable to contemplate 
instrumenting a volume of water or ice as large as $1\km^{3}$.  
\cite{detectors,DUMAND,KandSK,Baikal,Nestor,antares,Amanda,RMcKLR}.  
The ground array of the proposed Pierre Auger Cosmic Ray Observatory 
\cite{auger} would have an acceptance exceeding $1\km^{3}\:$sr of 
water for neutrino energies greater than $10^{17}\ev$ \cite{augeracc}.  
The Orbiting Wide-angle Light collectors project (\textsc{owl}) would 
place in Earth orbit a lens to study air showers initiated by 
$>10^{20}$-eV particles, including neutrinos \cite{owl}.

In this paper we present new calculations of the cross sections for 
charged-current and neutral current interactions of neutrinos with 
nucleons, updating our results of Ref.\ \cite{GQRS} (GQRS96) to take 
account of new information about the parton distributions within the 
nucleon \cite{pastis}.  In place of the CTEQ3--DIS parton 
distributions \cite{cteq} that we adopted as our nominal set in 
GQRS96, we base our new calculations on the CTEQ4--DIS parton 
distributions \cite{CTEQ4}.  The changes are modest, and only 
noticeable at the highest neutrino energies we consider: at 
$E_{\nu}=10^{21}\ev$, the new cross sections are about 25\% smaller 
than those of GQRS96.  We find that for neutrino energies up to 
$10^{16}\ev$, all the standard sets of parton distribution functions 
yield very similar cross sections.  At higher energies, the 
predictions rely on incompletely tested assumptions about the behavior 
of parton distributions at very small values of the momentum fraction 
$x$.  The resulting uncertainty reaches a factor $2^{\pm1}$ around 
$10^{20}\ev$.

We combine our new evaluations of the neutrino-nucleon cross sections 
with models for the flux of ultrahigh-energy (UHE) neutrinos to 
estimate event rates in neutrino observatories.  We consider the 
diffuse flux of neutrinos from AGNs and the flux of neutrinos that may 
accompany gamma-ray bursts, as well as neutrinos from cosmological 
sources such as the decay of topological defects formed in the early 
universe.  We evaluate rates for upward-going muons produced in or 
beneath large underwater and ice detectors, and we compute rates for 
contained neutrino interactions in a km$^{3}$ volume.  We have also 
estimated rates for the proposed ground array of the Pierre Auger 
Cosmic Ray Observatory.

The detection of upward-going muons from AGNs looks feasible in the 
next generation of underground experiments with effective areas on the 
order of $0.1\km^{2}$.  As the muon energy threshold increases above a 
few TeV, atmospheric neutrinos and muons become less important 
backgrounds.  Downward and air-shower event rates look promising for 
km$^{3}$ detectors, for a variety of models.

In the next Section, we review the calculation of the neutrino-nucleon 
charged-current and neutral-current cross sections, their 
sensitivities to parton distribution functions, and the resulting 
neutrino-nucleon interaction lengths.  We also give a brief account of 
neutrino-electron cross sections and interaction lengths.  
Ultrahigh-energy neutrino rates for a selection of flux models appear 
in \S\ref{sec:evts}.  Our summary and conclusions make up 
\S\ref{sec:conc}.

\section{Neutrino-nucleon interactions}\label{sec:nuN}
\subsection{Inclusive cross sections}\label{subsec:sigmanuN}
We calculate the inclusive cross section for the 
reaction
\begin{equation}
	\nu_\mu N \rightarrow \mu^- + \hbox{anything}  ,
\end{equation} 
where $N\equiv\displaystyle{\frac{n+p}{2}}$ is an 
isoscalar nucleon,
in the renormalization group-improved parton model. The differential cross 
section is written in terms of the Bjorken scaling variables 
$x = Q^2/2M\nu$  and $y = \nu/E_\nu$ as
\begin{equation}
	\frac{d^2\sigma}{dxdy} = \frac{2 G_F^2 ME_\nu}{\pi} \left( 
\frac{M_W^2}{Q^2 + M_W^2} \right)^{\!2} \left[xq(x,Q^2) + x 
\bar{q}(x,Q^2)(1-y)^2 \right] , \label{eqn:sigsig}
\end{equation} where $-Q^2$ is the invariant momentum transfer between 
the incident 
neutrino and outgoing muon, $\nu = E_\nu - E_\mu$ is the energy loss in 
the lab (target) frame, $M$ and $M_W$ are the nucleon and 
intermediate-boson masses, and $G_F = 1.16632 \times 10^{-5}~\rm{GeV}^{-2}$ is the Fermi 
constant. The quark distribution functions are
\begin{eqnarray}
q(x,Q^2) & = & \frac{u_v(x,Q^2)+d_v(x,Q^2)}{2} +
\frac{u_s(x,Q^2)+d_s(x,Q^2)}{2} \nonumber\\ & & + s_s(x,Q^2) + b_s(x,Q^2)  
 \\[12pt]
	\bar{q}(x,Q^2) & = & \frac{u_s(x,Q^2)+d_s(x,Q^2)}{2} + c_s(x,Q^2) + 
	t_s(x,Q^2),\nonumber 
\end{eqnarray}
where the subscripts $v$ and $s$ label valence and sea contributions, and 
$u$, $d$, $c$, $s$, $t$, $b$ denote the distributions for various quark 
flavors in a {\em proton}.  At the energies of interest for neutrino 
astronomy, perturbative QCD corrections to the cross section formula 
\eqn{eqn:sigsig} are insignificant.  In the DIS factorization scheme 
appropriate to the CTEQ4--DIS parton distributions, the terms 
proportional to $\alpha_{s}$ \cite{3elli} in the NLO cross section 
contribute only a few percent, so we omit them.

The $t\bar{t}$ sea is a negligible component of the nucleon over the 
$Q^{2}$-range relevant to neutrino-nucleon scattering; accordingly we 
neglect it.  At the energies of interest here, it is a sound 
kinematical simplification to treat charm and bottom quarks as 
massless.  However, the threshold suppression of the $b \rightarrow t$ 
transition must be taken into account.  We adopt the standard 
``slow-rescaling'' prescription \cite{xiscale}, with 
$m_{t}=175\gevcc$.  We have carried out numerical integrations using 
the adaptive Monte Carlo routine \textsc{vegas} \cite{vegas}, and 
Gaussian techniques.

The neutral-current cross section is of interest because it 
contributes to the attenuation of neutrinos as they pass through the 
Earth.  Neutral-current $\nu N$ interactions may also be significant 
backgrounds to the observation of the resonant formation process 
$\bar{\nu}_{e} \rightarrow W^{-}$.  If it becomes possible to 
measure neutral-current reactions and characterize the neutrino 
energy, Carena, \etal \cite{rpv} have shown that the neutral-current to 
charged-current ratio is an important discriminant of new physics.

Within the electroweak theory, a  calculation parallel to the one 
described above leads to the 
neutral-current cross section.  The differential cross section for the 
reaction $\nu_\mu N \rightarrow \nu_{\mu} + \hbox{anything}$ is given by
\begin{equation}
	\frac{d^2\sigma}{dxdy} = \frac{G_F^2 ME_\nu}{2\pi} \left( 
\frac{M_Z^2}{Q^2 + M_Z^2} \right)^{\!2} \left[xq^0(x,Q^2) + x 
\bar{q}^0(x,Q^2)(1-y)^2 \right] ,
\end{equation} 
where $M_Z$ is the mass of the neutral intermediate boson. 
The quantities involving parton distribution functions are
\begin{eqnarray}
	\lefteqn{q^0(x,Q^2) = \left[\frac{u_v(x,Q^2)+d_v(x,Q^2)}{2} +
\frac{u_s(x,Q^2)+d_s(x,Q^2)}{2}\right](L_u^2+L_d^2) } \nonumber \\ 
 & & \hspace{0.3in}+\;\left[\frac{u_s(x,Q^2)+d_s(x,Q^2)}{2}\right](R_u^2+R_d^2) + \\
 & & \hspace{0.3in}[s_s(x,Q^2) + b_s(x,Q^2)](L_d^2+R_d^2)
+[c_s(x,Q^2)+t_s(x,Q^2)](L_u^2+R_u^2)   
\nonumber \\[12pt]
	\lefteqn{\bar{q}^0(x,Q^2) = \left[\frac{u_v(x,Q^2)+d_v(x,Q^2)}{2} +
\frac{u_s(x,Q^2)+d_s(x,Q^2)}{2}\right](R_u^2+R_d^2) }  \nonumber \\
 & & \hspace{0.3in}+\;\left[\frac{u_s(x,Q^2)+d_s(x,Q^2)}{2}\right](L_u^2+L_d^2) + \\
 & & \hspace{0.3in}[s_s(x,Q^2) +
b_s(x,Q^2)](L_d^2+R_d^2) +[c_s(x,Q^2)+t_s(x,Q^2)](L_u^2+R_u^2), \nonumber
\end{eqnarray}
where the chiral couplings are
\begin{equation}
	\begin{array}{lcl}
		L_u = 1-\frac{4}{3}x_W & \phantom{WW} & L_d = -1 +\frac{2}{3}x_W  \\
		
		R_u = -\frac{4}{3}x_W &  & R_d = \frac{2}{3}x_W
	\end{array}
	\label{chicoups}
\end{equation}
and $x_W = \sin^2\theta_W$ is the weak mixing parameter. For numerical
calculations we have chosen $x_W = 0.226$, consistent with recent 
measurements \cite{schaile}.  Again the 
top-quark sea is negligible.

\begin{figure}[tbp]
	%\centering
		\centerline{\BoxedEPSF{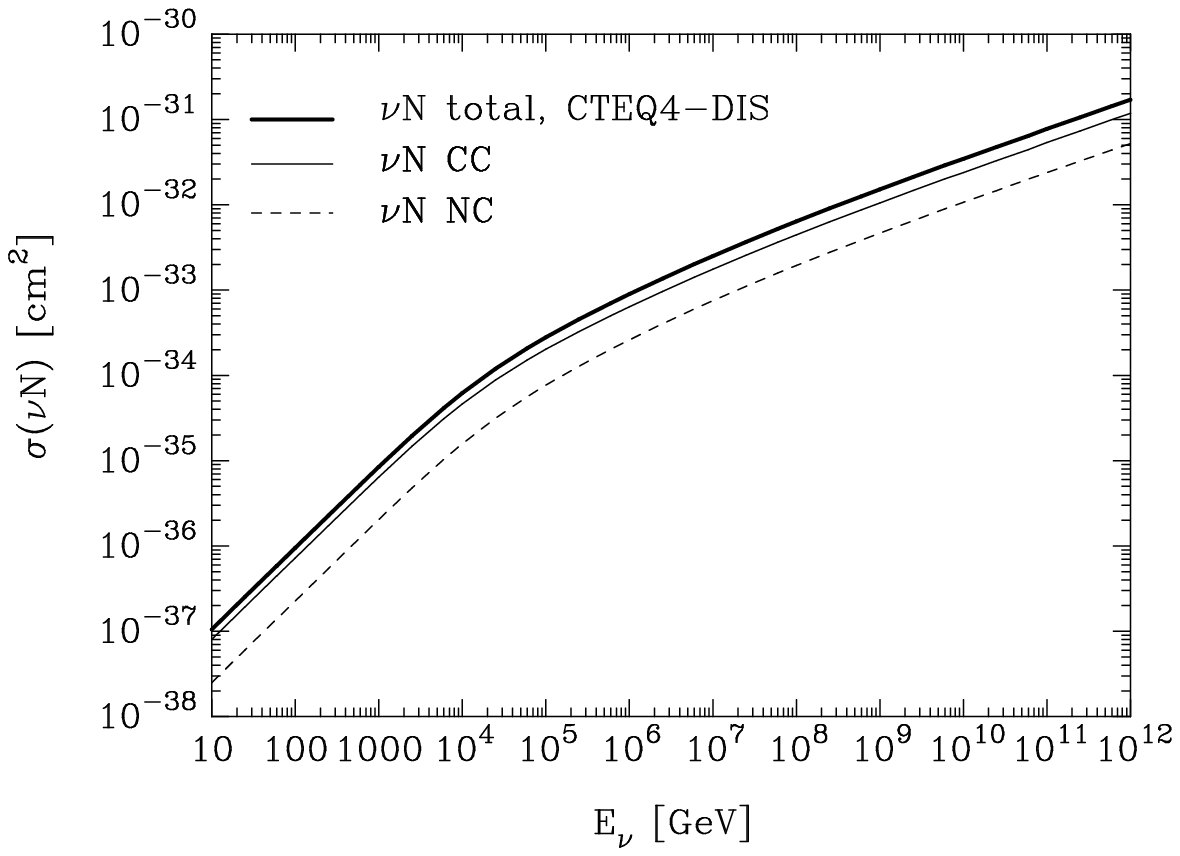  scaled 725}}
\caption{Cross sections for $\nu_{\ell} N$ interactions at high 
energies, according to the CTEQ4--DIS parton distributions: dashed 
line, $\sigma(\nu_{\ell} N \rightarrow \nu_{\ell}+\hbox{anything})$; 
thin line, $\sigma(\nu_{\ell} N \rightarrow 
\ell^{-}+\hbox{anything})$; thick line, total (charged-current plus 
neutral-current) cross section.}
	\label{fig:signuN4}
\end{figure}

We show the cross section for the charged-current reaction 
$\nu_\mu N \rightarrow \mu^- + \hbox{anything}$ as a function of the
neutrino energy $E_{\nu}$ in Figure \ref{fig:signuN4} (thin solid line).  
At low energies the charged-current cross section $\sigma_{\mathrm{CC}}$ 
rises linearly with $E_{\nu}$.  For energies exceeding about 
$10^{4}\gev$, the cross section is damped by the $W$-boson 
propagator.  We also show in Figure \ref{fig:signuN4} the neutral-current 
cross section $\sigma_{\mathrm{NC}}$ for the reaction 
$\nu_\mu N \rightarrow \nu_{\mu} + \hbox{anything}$ (dashed line), 
together with $\sigma_{\mathrm{tot}}$, the sum of charged-current and 
neutral-current cross sections (thick solid line).  For the range of 
neutrino energies of interest here, the charged-current results apply 
equally to the reaction $\nu_e N \rightarrow e^- + \hbox{anything}$.  
The neutral-current cross section for the reaction
$\nu_e N \rightarrow \nu_{e} + \hbox{anything}$ is identical to 
$\sigma_{\mathrm{NC}}$ depicted here.

\begin{figure}[tbp]
	%\centering
	\centerline{\BoxedEPSF{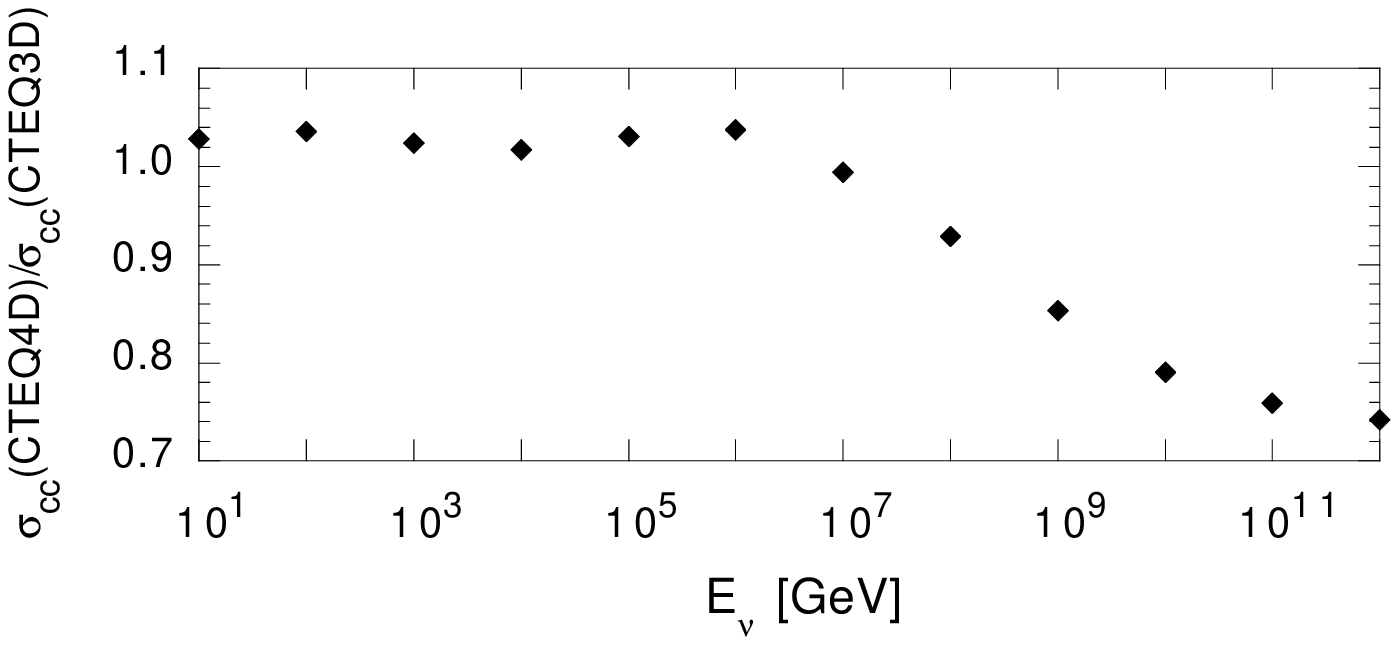  scaled 600}}
	\caption{Ratio of the charged-current cross section shown in Figure 
	{\protect\ref{fig:signuN4}}, calculated using the CTEQ4--DIS parton distributions, 
	to the charged-current cross section of Ref.~{\protect\cite{GQRS}} calculated 
	using the CTEQ3--DIS parton distributions.}
	\label{fig:cc4to3}
\end{figure}

The CTEQ4--DIS parton distributions are somewhat less singular as $x 
\rightarrow 0$ than the CTEQ3--DIS parton distributions we adopted as 
our nominal set in GQRS96.  Specifically, the sea-quark distributions 
of the CTEQ4 set behave as
\begin{equation}
	x q_{s}^{[\mathrm{CTEQ4}]}(x) \propto x^{-0.227}
	\label{eq:sing4}
\end{equation}
near $x=0$, whereas those of the CTEQ3 set behave as
\begin{equation}
	x q_{s}^{[\mathrm{CTEQ3}]}(x) \propto x^{-0.332}\; .
	\label{eq:sing3}
\end{equation}
The gentler singularity of the CTEQ4 distributions implies a smaller 
cross section at the highest energies, where the predominant 
contributions to the cross section come from very small values of 
$x$.  We show in Figure \ref{fig:cc4to3} the ratio of the 
charged-current cross sections calculated using the CTEQ4--DIS and 
CTEQ3--DIS parton distributions.  Up to $E_{\nu}\approx 10^{7}\gev$, 
the two evaluations agree within a few percent.  At still higher 
energies, the CTEQ4 cross section falls below the CTEQ3 cross 
section.  At the highest energy we consider, $E_{\nu}=10^{12}\gev$, 
the ratio is 0.74.  This is a small change.

\begin{figure}[tbp]
	%\centering
	\centerline{\BoxedEPSF{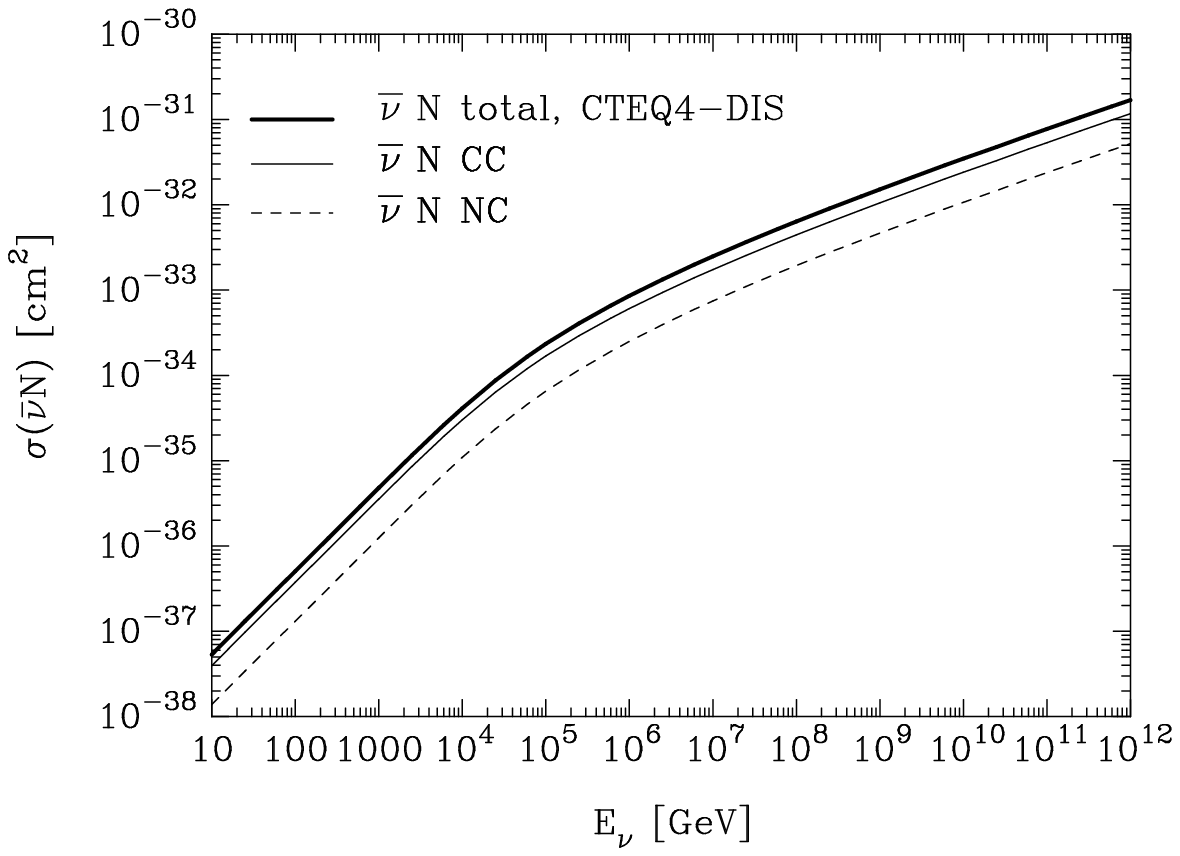  scaled 725}}
\caption{Cross sections for $\bar{\nu}_{\ell}N$ interactions at high 
energies, according to the CTEQ4--DIS parton distributions: dashed 
line, $\sigma(\bar{\nu}_{\ell}N \rightarrow 
\bar{\nu}_{\ell}+\hbox{anything})$; thin line, 
$\sigma(\bar{\nu}_{\ell} N \rightarrow \ell^{+}+\hbox{anything})$; 
thick line, total (charged-current plus neutral-current) cross 
section.}
	\label{fig:siganuN4}
\end{figure}

Similar calculations lead to the cross sections for $\bar{\nu}N$ 
scattering.  We show in Figure \ref{fig:siganuN4} the 
neutral-current (dashed line), charged-current (thin solid line), and 
total (thick solid line) $\bar{\nu}N$ cross sections.  At low energies, where the 
contributions of valence quarks predominate, the 
$\bar{\nu}N$ cross sections are smaller than the corresponding $\nu N$ 
cross sections, because of the familiar $(1-y)^{2}$ behavior of the 
$\bar{\nu}q$ cross sections.  Above $E_{\nu} \approx 10^{6}\gev$, the 
valence contribution is negligible and the $\nu N$ and $\bar{\nu}N$ 
cross sections become equal.

We collect in Tables \ref{tab:nuN4} and \ref{tab:anuN4} the 
charged-current, neutral-current, and total cross sections for $\nu N$ 
and $\bar{\nu}N$ interactions, respectively.  For the angular 
distributions, characterized by the mean inelasticity parameter 
$\langle y \rangle$, we refer to the CTEQ3 values we presented in 
Tables 1 and 2 of Ref.\ \cite{GQRS}.

For $10^{16}\ev \le E_{\nu} \le 
10^{21}\ev$, the CTEQ4--DIS cross sections 
are given within 10\% by
\begin{eqnarray}
	\sigma_{\mathrm{CC}}(\nu N) = 5.53 \times 10^{-36}\cm^{2}
	\left(\frac{E_{\nu}}{1\gev}\right)^{0.363} &  & \nonumber \\[6pt]
	\sigma_{\mathrm{NC}}(\nu N) = 2.31 \times 10^{-36}\cm^{2}
	\left(\frac{E_{\nu}}{1\gev}\right)^{0.363} &  & \nonumber \\[6pt]
	\sigma_{\mathrm{tot}}(\nu N) = 7.84 \times 10^{-36}\cm^{2}
	\left(\frac{E_{\nu}}{1\gev}\right)^{0.363} &  & \nonumber \\[12pt]
	\sigma_{\mathrm{CC}}(\bar{\nu} N) = 5.52 \times 10^{-36}\cm^{2}
	\left(\frac{E_{\nu}}{1\gev}\right)^{0.363} &  & \\[6pt]
	\sigma_{\mathrm{NC}}(\bar{\nu} N) = 2.29 \times 10^{-36}\cm^{2}
	\left(\frac{E_{\nu}}{1\gev}\right)^{0.363} &  & \nonumber \\[6pt]
	\sigma_{\mathrm{tot}}(\bar{\nu} N) = 7.80 \times 10^{-36}\cm^{2}
	\left(\frac{E_{\nu}}{1\gev}\right)^{0.363} &  & \nonumber .
\end{eqnarray}

\subsection{Variant parton distributions}\label{sub:HJ}
The CDF Collaborations's suggestion \cite{cdfjets} that the yield of 
jets in the reaction $\bar{p}p \rightarrow 
\hbox{jet}_{1}+\hbox{jet}_{2}+\hbox{anything}$ exceeds the rate 
expected in standard quantum chromodynamics has prompted a 
re-examination of the uncertainties of parton distributions at 
moderate and large values of $x$.  In particular, the CTEQ 
Collaboration has produced a variant of the CTEQ4 distributions, 
labelled CTEQ4--HJ, in which an enhanced gluon population at large 
values of $x$ raises the predicted two-jet inclusive cross section.  
The increased gluon density at large $x$ can affect the small-$x$ 
sea-quark distributions at large values of $Q^{2}$, so it is 
interesting to ask what difference the variant parton distributions 
would make for the UHE neutrino cross sections.  

\begin{figure}[tbp]
	%\centering
	\centerline{\BoxedEPSF{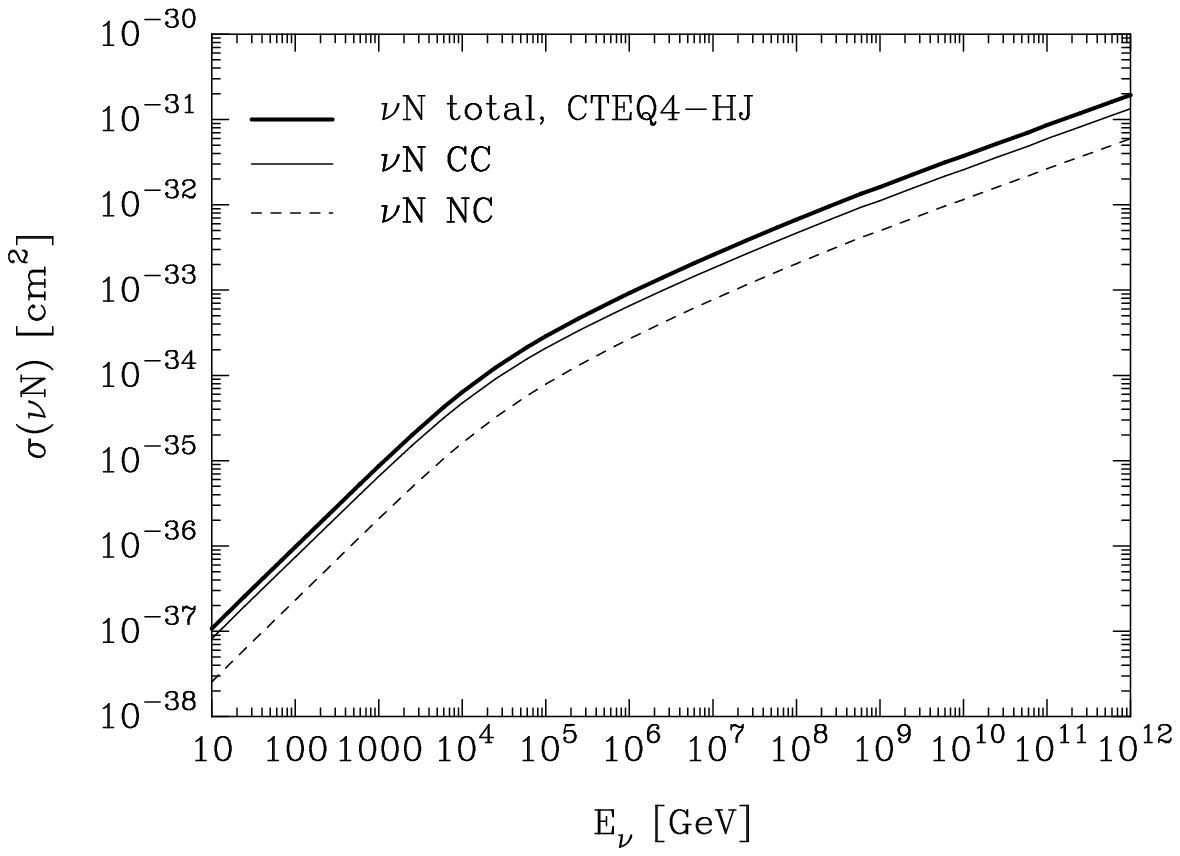  scaled 725}}
\caption{Cross sections for $\nu_{\ell}N$ interactions at high energies, 
according to the CTEQ4--HJ parton distributions: dashed line, 
$\sigma(\nu_{\ell} N \rightarrow \nu_{\ell}+\hbox{anything})$; thin 
line, $\sigma(\nu_{\ell} N \rightarrow \ell^{-}+\hbox{anything})$; 
thick line, total (charged-current plus neutral-current) cross 
section.}
	\label{fig:signuN4HJ}
\end{figure}

\begin{figure}[tbp]
	%\centering
	\centerline{\BoxedEPSF{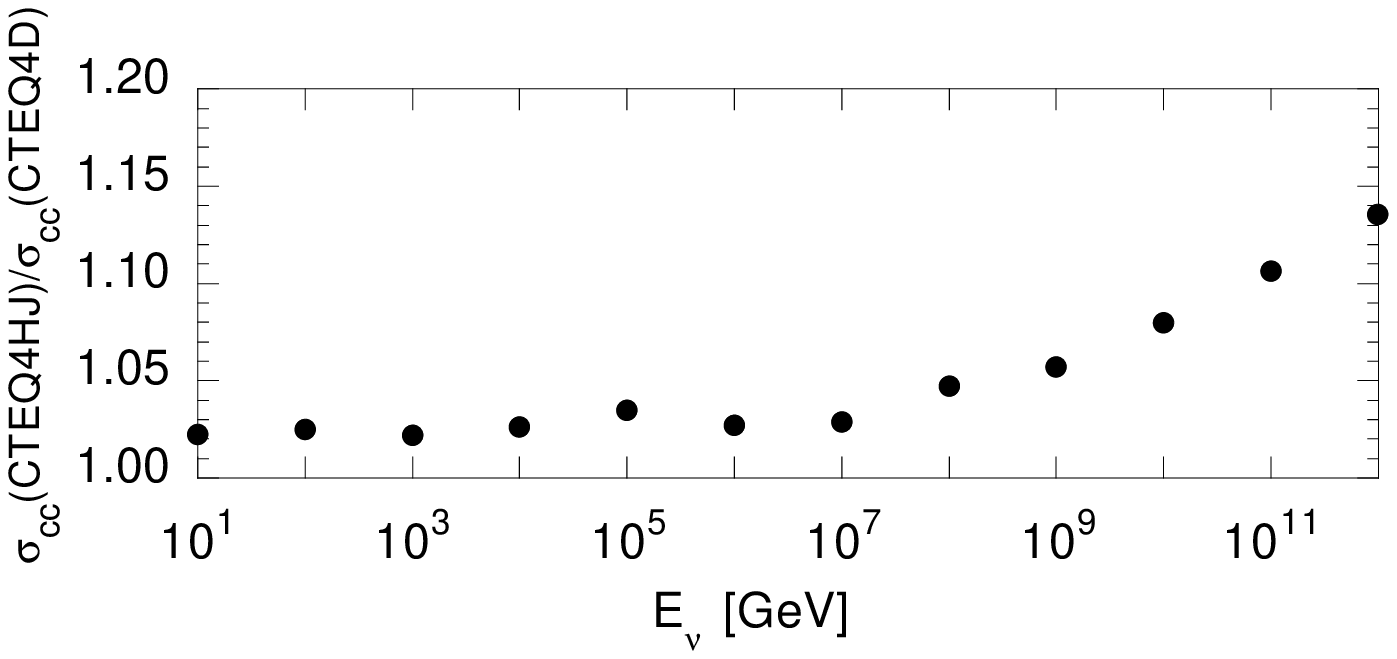  scaled 600}}
\caption{Ratio of the charged-current cross section shown in Figure 
{\protect\ref{fig:signuN4HJ}}, calculated using the CTEQ4--HJ parton 
distributions, to the charged-current cross section of Figure 
{\protect\ref{fig:signuN4}} calculated using the CTEQ4--DIS parton 
distributions.}
	\label{fig:ccHJto4}
\end{figure}

We show in Figure \ref{fig:signuN4HJ} the UHE $\nu N$ cross sections 
calculated with the CTEQ4--HJ parton distributions.  They are quite 
similar to those calculated with the standard CTEQ4--DIS 
distributions.  The ratio of CTEQ4--HJ to CTEQ4--DIS cross sections 
is displayed in Figure \ref{fig:ccHJto4}.  The difference is 
less than 5\% up to $10^{8}\gev$, and is smaller than 15\% at the 
highest energy we consider.  It is of no consequence for neutrino 
observatories.

\subsection{Assessment}\label{sub:assess}
How well is it possible to predict $\sigma(\nu_{\ell}N \rightarrow \ell + 
\hbox{anything})$ and $\sigma(\nu_{\ell}N \rightarrow \nu_{\ell} + 
\hbox{anything})$?  For $E_{\nu} \lesssim 10^{16}\ev$, all the standard 
sets of parton distributions, by which we mean those fitted to a vast 
universe of data, yield very similar cross 
sections\cite{GQRS,ghill,pz,Zhel}, within the standard electroweak 
theory.  For $E_{\nu} \gtrsim 10^{16}\ev$, 
cross sections are sensitive to the behavior of parton distributions 
at very small $x$, where there are no direct experimental constraints.
At these high energies, different \emph{assumptions} about $x 
\rightarrow 0$ behavior then lead to different cross sections.  
Judging from the most extreme variations we found in GQRS96, and from 
our present calculations, we conclude that at $10^{20}\ev$, the 
uncertainty reaches a factor of $2^{\pm 1}$.

New physics can, of course, modify the UHE cross sections.  The 
contributions of superpartners have been evaluated by Carena, \etal\ 
\cite{rpv}.  Doncheski and Robinett \cite{nuLQ} have investigated 
leptoquark excitation.  Bordes and collaborators \cite{dual} have 
speculated that new interactions might dramatically increase the 
UHE $\nu N$ cross sections, but Burdman, Halzen, and Gandhi 
have countered \cite{gustavo}  that unitarity limits the growth of the 
UHE cross section.

\subsection{Interaction lengths}\label{sec:lint}
The neutrino beams produced in accelerator laboratories are purified 
by passage through several kilometers (water equivalent) of material, 
which absorbs any accompanying particles without attenuating the 
neutrino flux.  The story is different at the ultrahigh energies of 
interest to neutrino astronomy.  The rise of the charged-current and 
neutral-current cross sections with energy is mirrored in the decrease 
of the (water-equivalent) interaction length,
\begin{equation}
	{\mathcal L}_{\mathrm{int}}= \frac{1}{\sigma_{\nu 
	N}(E_{\nu})N_{\mathrm{A}}}\;,
	\label{Lint}
\end{equation} 
where $N_{\mathrm{A}} = 6.022 \times 10^{23}\hbox{ mol}^{-1}=6.022 
\times 10^{23}\hbox{ cm}^{-3}$ (water equivalent) is Avogadro's 
number.  The energy dependence of the interaction lengths for 
neutrinos on nucleons is shown in Figure \ref{fig:lintnuN4}.  We show 
separately the interaction lengths for charged-current and 
neutral-current reactions, as well as the interaction length 
corresponding to the total (charged-current plus neutral-current) 
cross section.  The same information is shown for antineutrinos on 
nucleons in Figure \ref{fig:lintanuN4}.  Above about $10^{16}\ev$, the 
two sets of interaction lengths coincide.  These results apply equally 
to $\nu_{e}N$ (or $\bar{\nu}_{e}N$) collisions as to $\nu_{\mu}N$ (or 
$\bar{\nu}_{\mu}N$) collisions.

Over the energy range of interest for neutrino astronomy, the 
interactions of $\nu_{e}$, $\nu_{\mu}$, and $\bar{\nu}_{\mu}$ with 
electrons in the Earth can generally be neglected in comparison to 
interactions with nucleons.  The case of $\bar{\nu}_{e}e$ interactions 
is exceptional, because of the intermediate-boson resonance formed in 
the neighborhood of $E_{\nu}^{\mathrm{res}}=M_{W}^{2}/2m \approx 6.3 \times 
10^{15}\ev$.  The resonant reactions $\bar{\nu}_{e}e \rightarrow 
W^{-} \rightarrow \bar{\nu}_{\mu}\mu$ and $\bar{\nu}_{e}e \rightarrow 
W^{-} \rightarrow \hbox{hadrons}$ may offer a detectable signal.  
At resonance, the reaction $\bar{\nu}_{e}e \rightarrow W^{-} 
\rightarrow \hbox{anything}$ significantly attenuates a 
$\bar{\nu}_{e}$ beam propagating through the Earth.  The 
water-equivalent interaction lengths corresponding to the 
neutrino-electron cross sections computed in \cite{GQRS} are 
displayed in Figure \ref{fig:lintnue}.  These are evaluated as
\begin{equation}
	{\mathcal L}_{\mathrm{int}}^{(e)}= \frac{1}{\sigma_{\nu 
	e}(E_{\nu})(10/18)N_{\mathrm{A}}}\;,
		\label{elint}
\end{equation} where $(10/18)N_{\mathrm{A}}$ is the number of 
electrons in a mole of water.

We have reviewed current knowledge of the structure of the Earth in 
Ref.  \cite{GQRS}.  To good approximation, the Earth may be regarded 
as a spherically symmetric ball with a complex internal structure 
consisting of a dense inner and outer core and a lower mantle of 
medium density, covered by a transition zone, lid, crust, and oceans.  
A neutrino emerging from the nadir has traversed a column whose depth 
is 11 kilotonnes/cm$^2$, or $1.1\times 10^{10}\cmwe$.  The Earth's 
diameter exceeds the charged-current interaction length of neutrinos 
with energy greater than $40\tev$.  In the interval $2\times 
10^{6}\gev \lesssim E_{\nu} \lesssim 2 \times 10^{7}\gev$, resonant 
$\bar{\nu}_{e}e$ scattering adds dramatically to the attenuation of 
electron antineutrinos.  At resonance, the interaction length due to 
the reaction $\bar{\nu}_{e}e \rightarrow 
W^{-}\rightarrow\hbox{anything}$ is 6 tonnes/cm$^{2}$, or $6\times 
10^{6}\cmwe$, or 60 kmwe.  The resonance is effectively extinguished 
for neutrinos that traverse the Earth.  In the estimates of event 
rates that follow in \S \ref{sec:evts}, we take account of the 
effect of attenuation on upward-going neutrinos.

\section{Astrophysical neutrino fluxes: UHE event rates}
\label{sec:evts}
%\section{Event Rates}
Since the publication of the GQRS96 cross sections and event rates 
\cite{GQRS}, several new models of the diffuse neutrino flux from 
active galactic nuclei (AGNs) have appeared \cite{proth,karlm}.  
In addition, Waxman and Bahcall \cite{waxb} have argued that it might be 
possible to detect neutrinos associated with gamma-ray bursts.  New 
models have also been put forward for the production of neutrinos in 
the decays of generic heavy particles \cite{slsc} and in the collapse 
of topological defects \cite{branden}.

It is therefore timely to reconsider the event rates to be expected in 
large-volume detectors.  We focus on the production of upward-going 
muons in the charged-current reactions $(\nu_{\mu},\bar{\nu}_{\mu})N 
\rightarrow (\mu^{-},\mu^{+})+\hbox{anything}$.  Upward-going muons are free 
of background from the flux of muons produced by cosmic-ray 
interactions in the atmosphere.  It is in any case advantageous to 
site a detector beneath several kmwe to shield it from the (downward) 
rain of atmospheric muons \cite{lvd}.  Even at $3\kmwe$ underground, a 
detector still sees more than 200 vertical muons$\Flux$, though most of 
these muons are quite soft.  If we impose the requirement that 
$E_{\mu}^{\mathrm{min}}> (10^{3},10^{4},10^{5})\gev$, the flux is $(7, 
3\times 10^{-2}, 6\times 10^{-5})$ muons$\Flux$ \cite{mufldet}.  As the 
incident zenith angle of the atmospheric muons increases, the 
background flux decreases.  For horizontal incidence and below, the 
muon rate observed underground should be largely background-free.  
There is another important reason for looking down: The few-km range 
of UHE muons means that large-volume detectors can observe 
charged-current events that occur not only within the instrumented 
volume, but also in the rock or water underlying the detector.  
Accordingly, the effective volume of a detector may be considerably 
larger than the instrumented volume, for upward-going muons.
For energies above $40\tev$, the Earth's diameter exceeds the 
interaction length of neutrinos.  At these energies it is beneficial 
to look for events induced by downward and horizontal neutrino 
conversions to muons \cite{km}.  

\subsection{Sources of UHE Neutrinos}\label{sub:fluxmodels}
We display in Figure \ref{fig:fluxes} the neutrino fluxes calculated 
in a number of models.  Neutrinos produced by cosmic-ray interactions 
in Earth's atmosphere dominate over other neutrino sources at energies 
below a few TeV.  In this energy r\'{e}gime, the flux of atmospheric 
neutrinos is derived from the decay of charged pions and kaons.  The 
dotted curve in Figure \ref{fig:fluxes} shows the angle-averaged 
atmospheric (ATM) $\nu_{\mu}+\bar{\nu}_{\mu}$ flux calculated by Volkova 
\cite{volkova,otheratm}, which we parametrize as
\begin{equation}
\frac{dN_{\nu_{\mu}+\bar{\nu}_{\mu}}}{dE_{\nu}} = 7.8\times 10^{-11}
\left(\frac{E_\nu}{1\tev}\right)^{\!\!-3.6}\!\!\flux.
\end{equation}
The prompt neutrino flux from charm production in the atmosphere, a 
small additional component that appears above $E_{\nu}\approx 1\tev$, 
has been estimated recently by Pasquali, \etal\ \cite{prs}.

Active galactic nuclei (AGNs) are the most powerful radiation sources 
in the universe, with luminosities on the order of 
$10^{45\pm 3}\hbox{ erg/s}$.  They are cosmic accelerators powered by 
the gravitational energy of matter falling in upon a supermassive 
black hole.  Protons accelerated to very high energies within an AGN 
may interact with matter in the accretion disk, or with ultraviolet 
photons in the bright jets along the rotation axis.  Charged pions 
produced in the resulting $pp$ or $p\gamma$ collisions decay into 
muons and muon neutrinos.  The subsequent muon decays yield 
additional muon neutrinos and electron neutrinos.  Neutrino emission 
from AGNs may constitute the dominant diffuse flux at energies above a 
few TeV.

The solid lines in Figure \ref{fig:fluxes} show the 
$\nu_{\mu}+\bar{\nu}_{\mu}$ fluxes predicted in several contemporary 
models for neutrino production in AGNs: Protheroe's model \cite{proth} 
of neutrinos produced in $p\gamma$ interactions, denoted AGN-P96 
($p\gamma$); and Mannheim's model \cite{karlm} of neutrino 
production in $p\gamma$ interactions, denoted AGN-M95 ($p\gamma$).  
The neutrino fluxes (AGN-SS91) \cite{stecker} based on the pioneering 
work of Stecker and collaborators are still consistent with 
measurements.  We retain the Stecker-Salamon flux considered in GQRS96 
as a baseline to indicate the changes in event rates due to our new 
evaluation of the cross section.

A mechanism for gamma-ray bursts put forward by Waxman and Bahcall 
\cite{waxb} also yields UHE neutrinos.  The isotropic flux (GRB-WB) 
that results from a summation over sources is given by 
\begin{equation}
\frac{dN_{\nu_{\mu}+\bar{\nu}_{\mu}}}{dE_\nu}={\mathcal{N}}\left(\frac{E_\nu}{1\gev}\right)^{\!\!-n}\flux,
\end{equation}
with $({\mathcal{N}}=4.0\times 10^{-13},n=1)$ for $E_\nu<10^5\gev$, 
and $({\mathcal{N}}=4.0\times 10^{-8},n=2)$ for $E_{\nu}> 10^{5}\gev$.  
Sigl, \etal\ \cite{slsc} have explored a class of models of exotic 
heavy particle decays that ultimately lead to neutrinos in the final 
state.  These are referred to as top-down (TD-SLSC) models.  In 
particular, we consider the model in which the heavy $X$-particles 
have mass $M_X=2\times 10^{16}\gevcc$, which may arise from the 
collapse of networks of ordinary cosmic strings or from annihilations 
of magnetic monopoles.  An interesting feature of this model is that 
the highest energy cosmic rays are photons.  In the conventional 
topological defects model, the network of long strings loses its 
energy to the gravitation radiation.  Wichoski, \etal\ \cite{branden} 
have proposed a model in which particle production is the dominant 
channel through which energy is lost.  Even with the observational 
limits from Fr\'{e}jus and Fly's Eye as constraints, this non-scaling 
model (denoted TD-WMB) produces a much higher neutrino flux than the 
TD--SLSC model of Sigl, 
\etal, for the largest possible string mass density.
The fluxes predicted by these exotic sources are 
shown as dashed curves in Figure \ref{fig:fluxes}.

The Super-Kamiokande Collaboration has recently presented evidence 
that muon neutrinos produced in the atmosphere oscillate into tau 
neutrinos or sterile neutrinos during their passage through the 
Earth \cite{SKmix}.  At the neutrino energies of interest for the 
detection of extraterrestrial sources, $E_{\nu}\gtrsim 1\tev$, the 
probability for atmospheric neutrinos to oscillate \textit{en route} to the 
detector,
\begin{equation}
	P_{\nu_{\mu}\rightarrow \nu_{x}} \approx \sin^{2}\left(1.27 \times 
	10^{-6}\frac{\Delta m^{2}}{10^{-3}\ev^{2}}\cdot \frac{L}{1\km} 
	\cdot\frac{1\tev}{E_{\nu}}\right)\; ,
	\label{eq:mixprob}
\end{equation}
is less than $10^{-3}$ for a path length $L$ comparable to an Earth 
diameter ($\approx 13,000\km$), if the neutrino mass-squared 
difference $\Delta m^{2} \approx 2.2 \times 10^{-3}\ev^{2}$.  For UHE 
neutrinos from distant sources, the oscillation probability may 
become interestingly large.

\subsection{Neutrino--nucleon interactions}\label{sub:nuN}
The upward-muon event rate depends on the $\nu_{\mu}N$ cross section 
in two ways: through the interaction length that governs the 
attenuation of the neutrino flux due to interactions in the Earth, and 
through the probability that the neutrino converts to a muon energetic 
enough to arrive at the detector with $E_\mu$ larger than the 
threshold energy $E_\mu^{\rm min}$.

\begin{figure}[tbp]
	%\centering
	\centerline{\BoxedEPSF{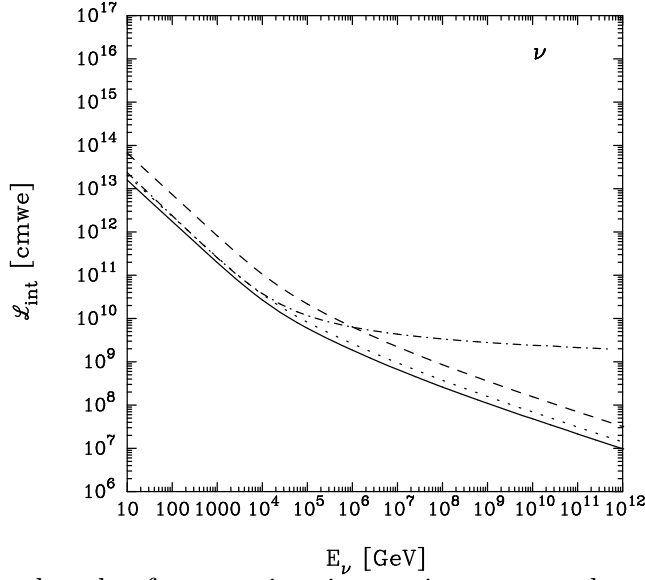  scaled 475}}
	\caption{Interaction lengths for neutrino interactions on nucleon 
	targets: dotted line, charged-current 
	interaction length; dashed line, neutral-current interaction length; 
	solid line, total interaction length, all computed with the CTEQ4--DIS 
	parton distributions.  The dot-dashed curve shows the charged-current 
	interaction length based on the EHLQ structure functions with $Q^{2}$ 
	held fixed at $Q_{0}^{2} = 5\gev^{2}$.}
	\label{fig:lintnuN4}
\end{figure}

\begin{figure}[tbp]
	%\centering
	\centerline{\BoxedEPSF{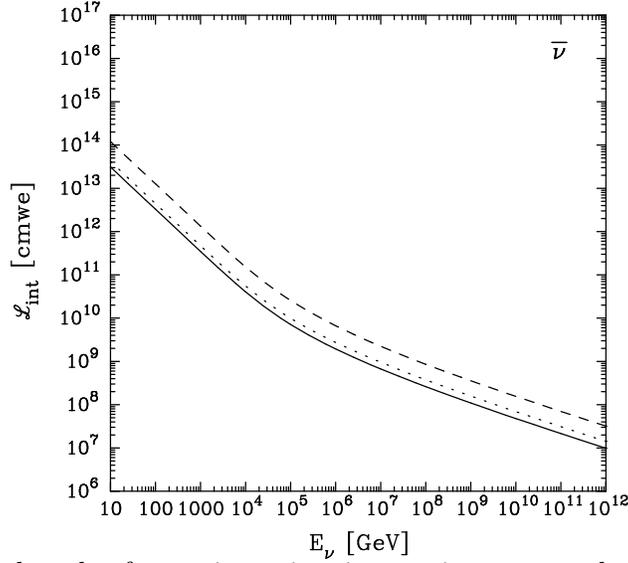  scaled 475}}
	\caption{Interaction lengths for antineutrino interactions on nucleon 
	targets: dotted line, charged-current 
	interaction length; dashed line, neutral-current interaction length; 
	solid line, total interaction length, all computed with the CTEQ4--DIS 
	parton distributions.}	\label{fig:lintanuN4}
\end{figure}

The probability that a muon produced in a charged-current interaction 
arrives in a detector with an energy above the muon energy threshold 
$E_\mu^{\mathrm min}$ is given by 
\begin{equation} 
P_\mu(E_\nu,E_\mu^{\rm min}) = N_{\mathrm{A}}\, \sigma_{\rm CC}(E_\nu) \langle 
R(E_\nu;E_\mu^{\rm min} )\rangle , \label{pmudef} 
\end{equation} 
where 
$\langle R(E_\nu;E_\mu^{\rm min})\rangle$ is the average range of a 
muon in rock \cite{ls} and $N_{\mathrm{A}}$ is Avogadro's number.  Although the 
Earth is transparent to low-energy neutrinos, an Earth diameter ($1.1 
\times 10^{10}\cmwe$) exceeds the interaction length of neutrinos with 
energies higher than about $40\tev$.  For the isotropic fluxes presented 
in \S \ref{sub:fluxmodels}, we represent the attenuation of neutrinos 
traversing the Earth by a shadow factor that is equivalent to the 
effective solid angle for upward muons, divided by $2\pi$ 
\cite{earth}:
\begin{equation}
S(E_\nu)={1\over 2\pi}\int_{-1}^{\:0} d\cos\theta\int d\phi
\exp \left[-z(\theta)/{\mathcal L}_{{\mathrm int}}(E_\nu) \right].
\label{Sdef}
\end{equation}
The column depth $z(\theta)$ is plotted in Figure 15 of Ref.  
\cite{GQRS}.  In our estimates of event rates, we choose the 
interaction length that corresponds to the total (charged-current plus 
neutral-current) cross section, which is shown for $\nu_{\mu}N$ and 
$\bar{\nu}_{\mu}N$ interactions as the solid curves in Figures 
\ref{fig:lintnuN4} and \ref{fig:lintanuN4}.  The shadow factor 
calculated using the CTEQ3--DIS parton distributions is shown in 
Figure 20 of Ref.\ \cite{GQRS}.

The rate at which upward-going muons can be observed in a detector 
with effective area $A$ is
\begin{equation}
{\mathrm Rate} = A \int_{E_{\mu}^{\mathrm{min}}}^{E^{\mathrm{max}}} dE_\nu\: P_\mu(E_\nu;E_\mu^{\rm min}) 
S(E_\nu){\frac{dN}{ dE_\nu}}. \label{rateqn}
\end{equation}
The integrals in \eqn{rateqn} are evaluated up to $E^{\mathrm 
max}=10^{11}\gev$ except for AGN-SS91, for which the data files extend 
only to $E^{\mathrm max}=10^{9.8}\gev$.

Let us consider for illustration a detector with effective area 
$A=0.1\km^{2}$.  We show in Tables \ref{tab:eth1} and \ref{tab:eth10} 
the annual event rates for upward-going muons with observed energies 
exceeding $1\tev$ and $10\tev$, respectively.  We tabulate rates for 
the full upward-going solid angle of $2\pi$, as well as for the 
detection of ``nearly horizontal'' muons with nadir angle $\theta$ 
between $60^{\circ}$ and $90^{\circ}$.  The predicted event rates, 
shown here for the CTEQ4--DIS parton distributions, are very 
similar for other modern parton distributions.

The AGN fluxes introduced in \S \ref{sub:fluxmodels} all yield 
significant rates for upward-going muon events, but the events induced 
by atmospheric neutrinos constitute an important background, 
especially for $E_{\mu}^{\mathrm{min}}=1\tev$.  The highest signal 
rates arise in the AGN-SS91 model.  At the lower muon 
energy threshold, $E_{\mu}^{\mathrm{min}}=1\tev$, this model yields 
rates about half the expected background from the interactions of 
atmospheric neutrinos.  The expected background exceeds the signal 
rates for other models with the muon energy threshold set at $1\tev$, 
but fades as the threshold is raised to $10\tev$.  When fully 
deployed, neutrino observatories of the current generation will have 
instrumented areas $A\approx 0.02\km^{2}$.  Although they should 
register a handful of AGN neutrinos, it will be difficult to 
distinguish them from the atmospheric background.  The energy 
dependence of the upward-going muon rate will be an important 
discriminant for separating atmospheric and extraterrestrial sources.

The GRB-WB flux yields a small number of upward-going muon events per 
year in a 0.1-km$^{2}$ detector.  To pick these out from the 
background, it will be essential to correlate them in time and 
position with gamma-ray bursts.  The TD-SLSC flux appears unobservable 
in a 0.1-km$^{2}$ detector.

The ratio of the ``nearly horizontal'' rate 
($60^{\circ}<\theta<90^\circ$) to the full upward rate characterizes 
the attenuation of the incident neutrino flux in the Earth.  The 
energy dependence of the neutrino interaction length shown in Figures 
\ref{fig:lintnuN4} and \ref{fig:lintanuN4} is reflected in the angular 
dependence of the shadow factor, for given neutrino energy: the 
greater the neutrino energy, the more the incident neutrino flux is 
attenuated in its passage through the Earth.  We recall that at 
$\theta=60^\circ$, the column depth is about 20\%\ of the vertical 
column depth \cite{GQRS}.  The ATM and AGN-M95 
fluxes fall rapidly with neutrino energy, so rates are dominated by 
neutrino energies near the muon energy threshold, and attenuation is 
not severe for $E_\mu^{\mathrm min}=1\tev$.  For these model fluxes, 
the upper half of the $2\pi$ solid angle for upward events contributes 
between 52\% and 58\% of the total number of upward-going events.
The other model fluxes decrease more slowly with neutrino energy.  
Accordingly, the effect of shadowing on the angular distribution is 
more pronounced.  For the AGN-SS91 and AGN-M95 fluxes, the ``nearly 
horizontal'' solid angle contributes 76\% and 87\% of the total.  The 
relatively stiff GRB-WB flux also entails significant shadowing: 67\% 
of the events come from the $1\pi$ solid angle just below the horizon.

The importance of shadowing increases as the muon energy threshold is 
raised.  For $E_\mu^{\mathrm min}=10\tev$, between 59\% and 89\% of 
the upward rate comes from the upper half of the upward solid angle.  

If we raise the muon energy threshold to $100\tev$, the atmospheric 
neutrino background is essentially eliminated.  At the same time, the 
projected signal rates, shown in Table \ref{tab:eth100} for a detector with 
$A=0.1\km^{2}$, reflect the influence of shadowing in the Earth in two 
ways.  First, the angular distributions favor shallow angles still 
more strongly.  Between 77\% and 94\% of the extraterrestrial neutrino 
signal comes from the ``nearly horizontal'' wedge.  Second, the shadow 
factor $S(E_{\nu})$ drops rapidly above $100\tev$.  It is equal to 
0.64, 0.34, 0.16, 0.071 for $E_{\nu} = 100\tev, \, 1\pev, \, 
10\pev, \hbox{ and }100\pev$.  Even so, the AGN-SS91 and 
AGN-P96 upward rates should be observable 
in a detector with an effective ares of $A=0.1\km^2$.

The penalty of shadowing overcomes the advantages of looking at 
upward-going muons as the muon-energy threshold is raised.  For very 
high thresholds, it will be necessary to observe downward-going muons 
produced by interactions within the instrumented volume.  We show in 
Table \ref{tab:hith} the annual downward event rates for $\nu_{\mu}N$ 
and $\bar{\nu}_{\mu}N$ charged-current interactions in a 1-km$^{3}$ 
volume of water.  The rates are encouraging, provided that 
downward-going events can be observed efficiently.

We discussed the effect of different choices for the parton 
distribution functions in GQRS96 for energy thresholds of $1\hbox{ and 
} 10\tev$.  We concluded that the upward muon event rate is 
insensitive to the choice of modern (post-HERA) parton distribution.  
We have evaluated the upward muon event rates for $E_\mu^{\rm 
min}=100\tev$ with the MRS D$\_^\prime$\cite{dmp} and 
CTEQ3-DLA\cite{dla} parton distributions.  The event rates are 
essentially equal to those shown in Table \ref{tab:eth100}.

We indicated in Section \ref{sub:assess} that all standard parton 
distribution functions yield very similar cross sections up to 
$E_{\nu} \approx 10^7\gev$, so only with a muon threshold 
$E_{\mu}^{\mathrm{min}} \gtrsim 10^{7}\gev$ might one distinguish 
between parton distribution functions.  The angular distribution of 
the upward muons is a measure of attenuation as a function of column 
depth, and thus of the cross section.  We show in Figure 
\ref{fig:atten} the differential shadow factor
$(1/2\pi)\: dS/d\cos\theta$ for three neutrino energies and for three 
sets of parton distributions.  At $E_{\nu}=10^{3}\gev$, there is no 
appreciable shadowing at any angle.  By $E_{\nu}=10^{6}\gev$, 
$S(E_{\nu})\approx0.3$, and the nearly vertical (upward-going) events 
are depleted in comparison to nearly horizontal events.  However, the 
angular distribution of events will be quite similar for our nominal 
set of parton distributions (CTEQ4--DIS) and for the CTEQ3--DLA and 
MRS-D$\_^\prime$ distributions.  At still higher energies, events will 
be observed only near the horizon.  The angular distribution is 
steepest for the MRS-D$\_^\prime$ distributions, which yield the 
largest $\nu N$ cross sections, and shallowest for the CTEQ3--DLA 
distributions, which yield the smallest $\nu N$ cross sections.  Given 
the low rates we anticipate for $E_{\nu}\gtrsim 10^{8}\gev$, a 
discriminating measurement would demand a prohibitively large 
instrumented volume.

\subsection{Neutrino-electron scattering}\label{sub:nue}
Observations of electron neutrino interactions at higher energies and 
large target volumes may yield insights into the ultrahigh energy 
neutrino flux and the high energy $\nu N$ cross section.  Generically, 
the $\nu_{e}+\bar{\nu}_{e}$ flux from the $\pi \rightarrow \mu 
\rightarrow e$ decay chain is one-half the $\nu_{\mu}+\bar{\nu}_{\mu}$ 
flux.  For the rate estimates presented here, we use the fluxes of 
Figure \ref{fig:fluxes} multiplied by 0.5.  As a cautionary note, we 
call attention to the contention of Rachen and Meszaros \cite{rachen} 
that muon cooling within astrophysical sources may reduce the UHE 
$\nu_{\mu}+\bar{\nu}_{\mu}$ flux by a factor of two and effectively 
eliminate the UHE $\nu_{e}+\bar{\nu}_{e}$ flux.  In this subsection, 
we shall first evaluate rates for resonant $W$ production.  Then we 
consider the possibility of observing nearly horizontal air showers 
induced by neutrino interactions in the atmosphere.

\begin{figure}[tbp]
	\centerline{\BoxedEPSF{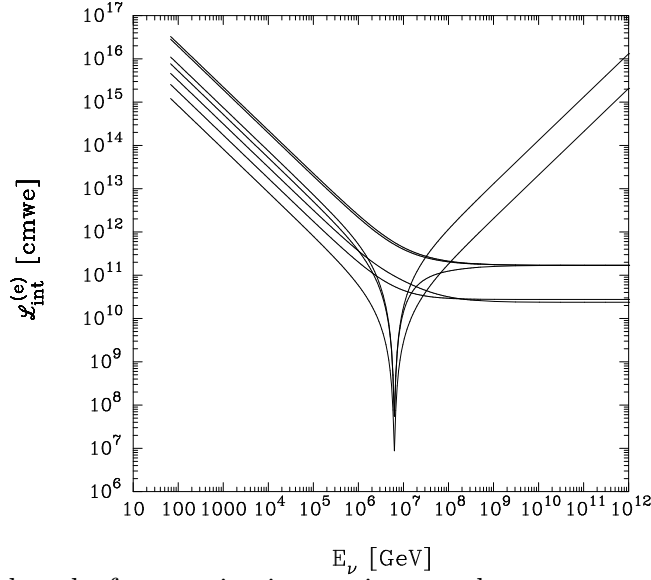  scaled 950}}
	\caption{Interaction lengths for neutrino interactions on electron 
	targets.  At low energies, from smallest to largest interaction length, 
	the processes are (i) $\bar{\nu}_{e}e \rightarrow \hbox{ hadrons}$, 
	(ii) $\nu_{\mu}e \rightarrow \mu\nu_{e}$, (iii) $\nu_{e}e \rightarrow 
	\nu_{e}e$, (iv) $\bar{\nu}_{e}e \rightarrow \bar{\nu}_{\mu}\mu$, (v) 
	$\bar{\nu}_{e}e \rightarrow \bar{\nu}_{e}e$, (vi) $\nu_{\mu}e 
	\rightarrow \nu_{\mu}e$, (vii) $\bar{\nu}_{\mu}e \rightarrow 
	\bar{\nu}_{\mu}e$.  [From Ref.~{\protect\cite{GQRS}}]} 
	\label{fig:lintnue}
\end{figure}

\begin{figure}[tbp]
\centerline{\BoxedEPSF{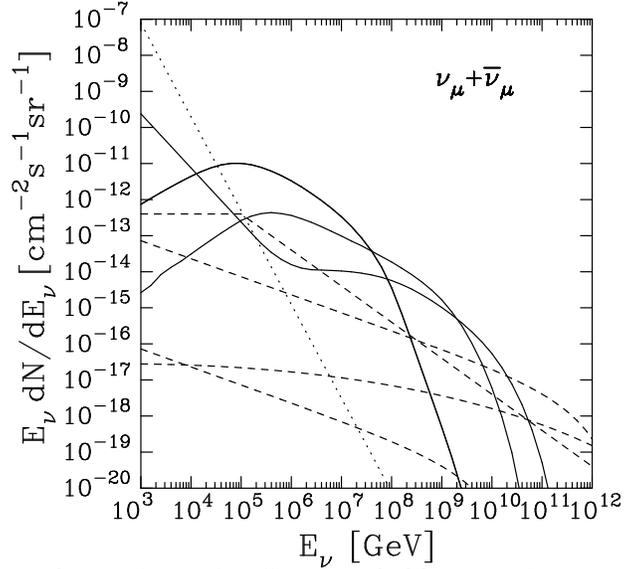  scaled 450}}
\caption{Muon neutrino plus antineutrino fluxes scaled by neutrino 
energy at the Earth's surface.  Solid lines represent AGN fluxes.  In 
decreasing magnitude at $E_\nu=10^3$ GeV, they are AGN-M95, AGN-SS91 
scaled by 0.3, and AGN-P96 ($p\gamma$).  The dashed lines, in the same 
order, represent the GRB-WB, TD-WMB12, TD-WMB16, and TD-SLSC fluxes.  
The dotted line is the angle-averaged atmospheric (ATM) neutrino 
flux.}
\label{fig:fluxes}
\end{figure}

\begin{figure}[tbp]
\centerline{\BoxedEPSF{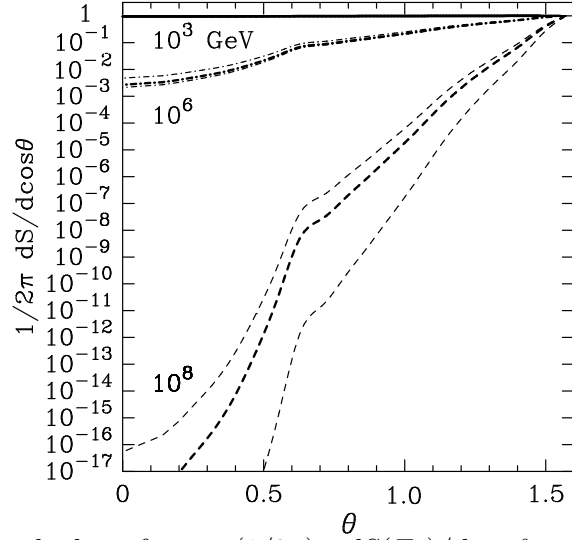  scaled 450}}
\caption{Differential shadow factor 
$(1/2\pi)\;dS(E_{\nu})/d\cos\theta$ \textit{versus} nadir angle 
$\theta$, for $E_\nu=10^3\gev$ (solid line), $10^6\gev$ (dot-dashed 
lines), and $10^8\gev$ (dashed lines).  For each neutrino energy, the 
thick line correspondes to the CTEQ4--DIS parton distributions; the 
upper and lower satellite lines correspond to the CTEQ3--DLA and 
MRS-D$\_^\prime$ parton distributions, respectively.}
\label{fig:atten}
\end{figure}

Resonant $\bar{\nu}_e e \rightarrow W^-$ 
production occurs for $E_{\bar{\nu}}\simeq 6.3\pev$. In an effective volume
$V_{\rm eff}$, the contained event rate is
\begin{equation}
{\mathrm{Rate}}={10\over 18}\, N_{\mathrm{A}}\, V_{\rm eff}\ \int_{(M_W-2\Gamma_W)^2/2m}^
{(M_W+2\Gamma_W)^2/2m} \!\!\!\!\!\!\!\!\!dE_{\bar{\nu}_e}\, \sigma_{\bar{\nu}_e e}
(E_{\bar{\nu}_e}) \, {dN_{\bar{\nu}_e}\over dE_{\bar{\nu}_e}}\ .
\end{equation}
Downward resonant $\bar{\nu}_e e\rightarrow W^-$ rates are shown in 
Table \ref{tab:glashow} for an effective volume of 1 km$^3$.  To 
assess potential backgrounds to the detection of $\bar{\nu}_{e}e 
\rightarrow W^{-} \rightarrow \mu\bar{\nu}_{\mu}$ and 
$\bar{\nu}_{e}e\rightarrow W^{-} \rightarrow \hbox{hadrons}$, we also 
show the downward (and upward) $(\nu_\mu,\, \bar{\nu}_\mu)N$ 
charged-current and neutral-current events that occur for 
$E_\nu>3\pev$.  The $\nu_e N$ downward and upward interaction rates 
and the $\bar{\nu}_e N$ downward interaction rates may be obtained 
from the $(\nu_\mu, \bar{\nu}_\mu)N$ rates by scaling the incident 
fluxes.  The $\bar{\nu}_e N$ upward interaction rates are reduced by 
the short $\bar{\nu}_e e$ interaction length near resonance.  All of 
these backgrounds to the identification of $\bar{\nu}_e e \rightarrow 
W^- \rightarrow \bar{\nu}_\mu \mu \hbox{ or hadrons}$ are themselves 
evidence for extraterrestrial neutrinos.  We conclude, as in GQRS96, 
that resonant $W^-$ production will be difficult to extract from the 
neutrino-nucleon interaction background.  The short $\bar{\nu}_{e}$ 
interaction length at resonance (\cf Figure \ref{fig:lintnue}) means 
that the flux of electron antineutrinos is extinguished for neutrinos 
traversing the Earth.

Let us now take up the observability of neutrino interactions in the 
atmosphere.  A neutrino normally incident on a surface detector passes 
through a column of density of $1\,033\cmwe$, while a neutrino 
arriving along the horizon encounters a column of about 
$36\,000\cmwe$.  Both amounts of matter are orders of magnitude 
smaller than the neutrino interaction lengths summarized in Figures
\ref{fig:lintnuN4}--\ref{fig:lintnue}, so the atmosphere is essentially 
transparent to neutrinos.  

However, the horizontal path length low in the atmosphere is not tiny 
compared with the depth available for the production of contained 
events in a water or ice \v{C}erenkov detector \cite{GQRSfig16}, so it 
is worth asking what capabilities a large-area air-shower array might 
have for the study of UHE neutrino interactions.  The proposed Pierre 
Auger Cosmic Ray Observatory \cite{auger}, which would consist of an 
array of water \v{Cerenkov} tanks dispersed over a large land area, is 
designed to detect showers of particles produced in the atmosphere.  
Proton- and photon-induced showers are typically produced high in the 
atmosphere.  Nearly horizontal events with shower maxima near the 
surface array are more likely to arise from neutrino interactions than 
from $p$-Air or $\gamma$-Air collisions \cite{bpvz}.

\begin{figure}[tbp]
\centerline{\BoxedEPSF{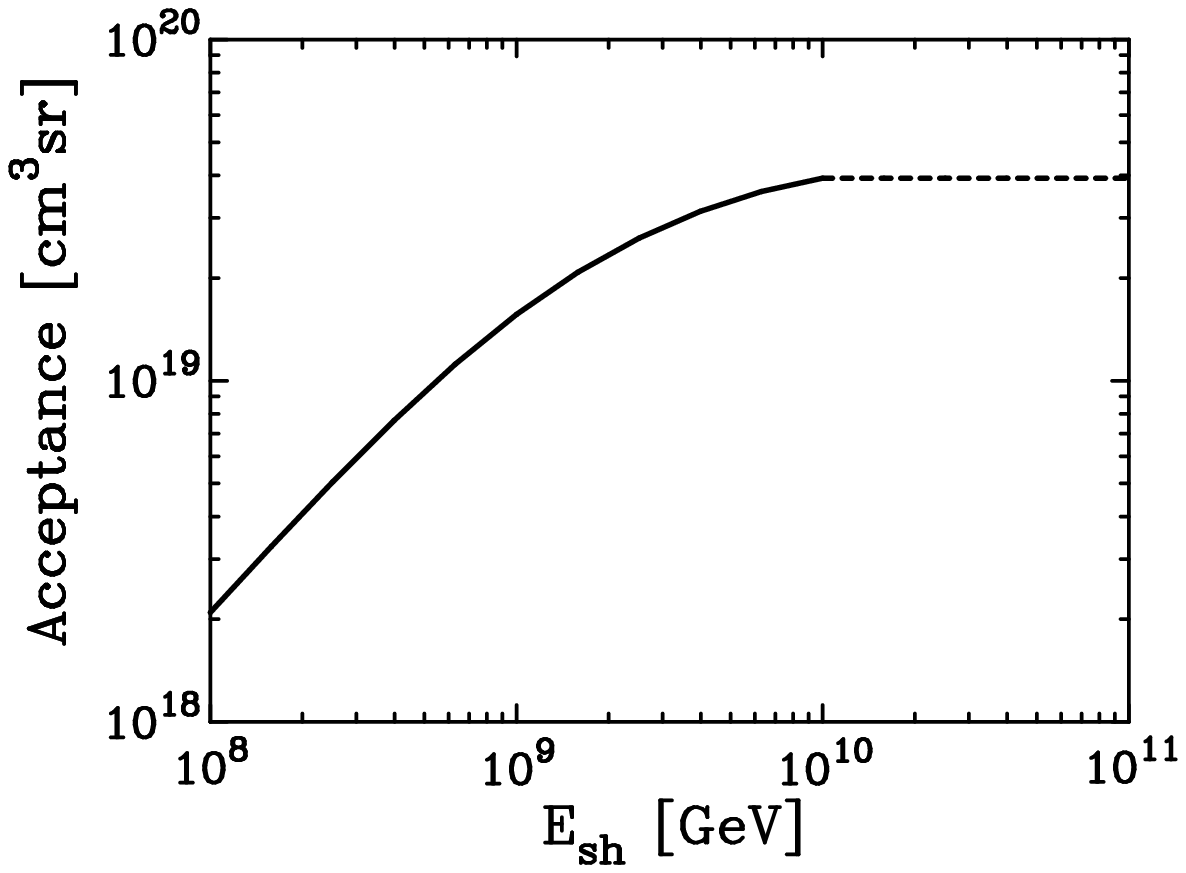  scaled 675}}
\caption{Acceptance versus shower energy for 
nearly horizontal air showers in the proposed Pierre Auger Cosmic Ray 
Observatory.  The solid line for $E_{\mathrm{sh}}\le 10^{10}\gev$ 
shows the acceptance evaluated by Billoir {\protect\cite{augeracc}}; 
the dashed line is our projection to higher energies.}
\label{fig:billoir}
\end{figure}

The acceptance ${\cal A}$ of the Auger ground array, which has 
dimensions of volume times solid angle, has been evaluated by several 
authors \cite{augeracc,augernu2,augernu}.  We adopt the Billoir's 
estimate \cite{augeracc}, as shown in Figure \ref{fig:billoir}, to 
compute the event rate
\begin{equation}
{\mathrm{Rate}} = N_{\!\mathrm{A}}\,\rho_{\rm air}
 \int_{E_{\mathrm{th}}}^{E^{\rm max}} \!\!\!\!\!\!\!\!dE_{\mathrm{sh}} 
 \int_0^1\!\!\!\!dy {dN_\nu\over
dE_\nu}{d\sigma_{\nu N}\over dy}(E_\nu,y){\cal A}(E_{\mathrm{sh}})\ .
\end{equation}
The values of $E^{\mathrm max}$ are the same as for our calculation 
of upward muon rates in \S\ref{sub:nuN}.  For 
$(\nu_{e}+\bar{\nu}_{e})N$ charged-current interactions, we take the 
shower energy to be the sum of hadronic and electromagnetic 
energies, $E_{\mathrm{sh}}=E_{\nu}$.  For 
$(\nu_{\mu}+\bar{\nu}_{\mu})N$ charged-current interactions and for 
neutral-current interactions, we take the shower energy to be the 
hadronic energy, $E_{\mathrm{sh}}=yE_{\nu}$.  The resulting event 
rates, calculated using our canonical CTEQ4--DIS parton distributions, 
are shown in Table \ref{tab:PAnue} for $(\nu_{e}+\bar{\nu}_{e})N$ 
neutral-current interactions and for $(\nu_{\mu}+\bar{\nu}_{\mu})N$ 
charged-current interactions, for two shower thresholds.  The 
$(\nu_{\mu}+\bar{\nu}_{\mu})N$ neutral-current rates are twice those 
shown for the $(\nu_{e}+\bar{\nu}_{e})N$ case.  In Table \ref{tab:auger} 
we show our evaluation of the $(\nu_{e}+\bar{\nu}_{e})N$ 
charged-current rates for three different sets of parton 
distributions.

The $(\nu_e+\bar{\nu}_e)N$ neutral-current event rates are typically 
less than 15\% of the corresponding charged-current rates, reflecting 
a combination of smaller cross sections and a falling flux.  The 
$(\nu_{\mu}+\bar{\nu}_{\mu})N$ charged-current rates are a factor of 
$\sim 4$ larger than the $(\nu_e+\bar{\nu}_e)N$ neutral-current rates, 
because of the larger flux and larger cross section.  The inelasticity 
parameters $\langle y \rangle$ are approximately equal for 
neutral-current and charged-current interactions.

The largest rates for neutrino-induced horizontal air showers arise 
from $(\nu_e+\bar{\nu}_e)N$ charged-current interactions, for which 
$E_{\mathrm{sh}} \approx E_{\nu}$.  In one 
year, a few to tens of horizontal $(\nu_e+\bar{\nu}_e)N \rightarrow 
e^{\mp}+\hbox{anything}$ events may be observed in the Auger 
detectors, assuming the modern estimates of AGN neutrino fluxes.  The 
AGN-SS91, GRB-WB, TD-SLSC, and TD-WMB16 fluxes yield fractions of an event 
per year, while the TD-WMB12 flux yields an event or two.

Given the high thresholds that must be set for detection, the 
expected event rates are dependent on the choice of parton
distribution functions. The D$\_^\prime$ rates are approximately
a factor of two larger than the CTEQ3-DLA rates. 
If the absolute normalization and energy behavior of the AGN fluxes
could be established in underground experiments at lower energies,
the Auger experiment might suggest distinctions among the various
high-energy extrapolations of the cross sections.

\section{Summary and outlook}\label{sec:conc}
We have presented new calculations of the cross sections for 
neutrino-nucleon charged-current and neutral-current interactions.  
The new cross sections are at most 25\% smaller than those of GQRS96, 
with the deviation largest at the highest energy considered here, 
$10^{21}\ev$.  By varying the extrapolations of the small-$x$ behavior 
of the parton distribution functions, we find that the uncertainty in 
the $\nu N$ cross section is at most a factor of $2^{\pm 1}$ at the 
highest energies.  All modern sets of parton distribution functions 
give comparable cross sections for energies up to $10^{16}\ev$.

We have estimated event rates for several energy thresholds and 
detection methods, using a variety of models for the neutrino fluxes 
from AGNs, gamma-ray bursters, topological defects, and cosmic-ray 
interactions in the atmosphere.  In $\nu_\mu N \rightarrow \mu X$ 
interactions, requiring a muon energy above 10 TeV reduces the 
atmospheric background enough to permit the observation of 
upward-going muons for the AGN-SS91 and AGN-P96 fluxes.  These models 
yield tens to hundreds of events per year for detectors of $0.1\km^2$ 
effective area.  The GRB-WB flux emerges at a higher threshold, but 
suffers from a small event rate.

Event rates for downward muons above 
$100\tev$ from neutrinos are substantial in $1\km^{3}$, except for the 
TD models.  Resonant $W$ boson production will be difficult to 
distinguish from the $\nu N$ interaction background.  For the Pierre 
Auger Cosmic Ray Observatory, the most promising rates arise from 
$(\nu_e ,\ \bar{\nu}_e) N$ charged-current interactions in the AGN-M95 
and AGN-P96 models.  By combining measurements of the upward-going 
muon rate at lower energies with air-shower studies at the highest 
energies, it may be possible to distinguish among alternative 
high-energy extrapolations of the $\nu N$ cross section.

The origins of the highest energy cosmic rays are not well understood, 
but cosmic rays should be accompanied by very high energy neutrinos in 
all models.  The absolute normalization and energy dependence of the 
fluxes vary from model to model.  Neutrino telescopes ultimately will 
probe extraterrestrial accelerator sources.  We expect that detectors 
with effective areas on the order of $0.1\km^2$ will yield significant 
clues to aid in our understanding of physics to the 10$^{20}$-eV 
energy scale.

\acknowledgments
RG is grateful for the hospitality of the Department of Physics at the 
University of Arizona and of the Fermilab Theoretical Physics 
Department.  Fermilab is operated by Universities Research 
Association, Inc., under contract DE-AC02-76CHO3000 with the United 
States Department of Energy.  CQ thanks the Department of Physics at 
Princeton University for warm hospitality.  MHR acknowledges the 
hospitality of the CERN Theory Division.  The research of MHR at the 
University of Iowa is supported in part by National Science Foundation 
Grant PHY~95-07688.  The research of IS at the University of Arizona 
is supported in part by the United States Department of Energy under 
contracts DE-FG02-85ER40213 and DE-FG03-93ER40792.  CQ, MHR, and IS 
acknowledge the hospitality of the Aspen Center for Physics.

\newpage
\narrowtext
\begin{table}
\caption{Charged-current and neutral-current cross sections and their 
sum for $\nu N$ interactions according to the 
CTEQ4--DIS distributions.}

\begin{center}
	\begin{tabular}{cccc}
		%\hline
		$E_{\nu}\hbox{ [GeV]}$ & $\sigma_{\mathrm{CC}}$ [cm$^2$] & 
		$\sigma_{\mathrm{NC}}$ [cm$^2$] & $\sigma_{\mathrm{tot}}$ [cm$^2$] \\
		\hline
$1.0\times 10^{1}$   &  $0.7988\times 10^{-37}$   &  $0.2492\times 10^{-37}$   &  $0.1048\times 10^{-36}$ \\
$2.5\times 10^{1}$   &  $0.1932\times 10^{-36}$   &  $0.6033\times 10^{-37}$   &  $0.2535\times 10^{-36}$ \\
$6.0\times 10^{1}$   &  $0.4450\times 10^{-36}$   &  $0.1391\times 10^{-36}$   &  $0.5841\times 10^{-36}$ \\
$1.0\times 10^{2}$   &  $0.7221\times 10^{-36}$   &  $0.2261\times 10^{-36}$   &  $0.9482\times 10^{-36}$ \\
$2.5\times 10^{2}$   &  $0.1728\times 10^{-35}$   &  $0.5430\times 10^{-36}$   &  $0.2271\times 10^{-35}$ \\
$6.0\times 10^{2}$   &  $0.3964\times 10^{-35}$   &  $0.1255\times 10^{-35}$   &  $0.5219\times 10^{-35}$ \\
$1.0\times 10^{3}$   &  $0.6399\times 10^{-35}$   &  $0.2039\times 10^{-35}$   &  $0.8438\times 10^{-35}$ \\
$2.5\times 10^{3}$   &  $0.1472\times 10^{-34}$   &  $0.4781\times 10^{-35}$   &  $0.1950\times 10^{-34}$ \\
$6.0\times 10^{3}$   &  $0.3096\times 10^{-34}$   &  $0.1035\times 10^{-34}$   &  $0.4131\times 10^{-34}$ \\
$1.0\times 10^{4}$   &  $0.4617\times 10^{-34}$   &  $0.1575\times 10^{-34}$   &  $0.6192\times 10^{-34}$ \\
$2.5\times 10^{4}$   &  $0.8824\times 10^{-34}$   &  $0.3139\times 10^{-34}$   &  $0.1196\times 10^{-33}$ \\
$6.0\times 10^{4}$   &  $0.1514\times 10^{-33}$   &  $0.5615\times 10^{-34}$   &  $0.2076\times 10^{-33}$ \\
$1.0\times 10^{5}$   &  $0.2022\times 10^{-33}$   &  $0.7667\times 10^{-34}$   &  $0.2789\times 10^{-33}$ \\
$2.5\times 10^{5}$   &  $0.3255\times 10^{-33}$   &  $0.1280\times 10^{-33}$   &  $0.4535\times 10^{-33}$ \\
$6.0\times 10^{5}$   &  $0.4985\times 10^{-33}$   &  $0.2017\times 10^{-33}$   &  $0.7002\times 10^{-33}$ \\
$1.0\times 10^{6}$   &  $0.6342\times 10^{-33}$   &  $0.2600\times 10^{-33}$   &  $0.8942\times 10^{-33}$ \\
$2.5\times 10^{6}$   &  $0.9601\times 10^{-33}$   &  $0.4018\times 10^{-33}$   &  $0.1362\times 10^{-32}$ \\
$6.0\times 10^{6}$   &  $0.1412\times 10^{-32}$   &  $0.6001\times 10^{-33}$   &  $0.2012\times 10^{-32}$ \\
$1.0\times 10^{7}$   &  $0.1749\times 10^{-32}$   &  $0.7482\times 10^{-33}$   &  $0.2497\times 10^{-32}$ \\
$2.5\times 10^{7}$   &  $0.2554\times 10^{-32}$   &  $0.1104\times 10^{-32}$   &  $0.3658\times 10^{-32}$ \\
$6.0\times 10^{7}$   &  $0.3630\times 10^{-32}$   &  $0.1581\times 10^{-32}$   &  $0.5211\times 10^{-32}$ \\
$1.0\times 10^{8}$   &  $0.4436\times 10^{-32}$   &  $0.1939\times 10^{-32}$   &  $0.6375\times 10^{-32}$ \\
$2.5\times 10^{8}$   &  $0.6283\times 10^{-32}$   &  $0.2763\times 10^{-32}$   &  $0.9046\times 10^{-32}$ \\
$6.0\times 10^{8}$   &  $0.8699\times 10^{-32}$   &  $0.3837\times 10^{-32}$   &  $0.1254\times 10^{-31}$ \\
$1.0\times 10^{9}$   &  $0.1049\times 10^{-31}$   &  $0.4641\times 10^{-32}$   &  $0.1513\times 10^{-31}$ \\
$2.5\times 10^{9}$   &  $0.1466\times 10^{-31}$   &  $0.6490\times 10^{-32}$   &  $0.2115\times 10^{-31}$ \\
$6.0\times 10^{9}$   &  $0.2010\times 10^{-31}$   &  $0.8931\times 10^{-32}$   &  $0.2903\times 10^{-31}$ \\
$1.0\times 10^{10}$   &  $0.2379\times 10^{-31}$   &  $0.1066\times 10^{-31}$   &  $0.3445\times 10^{-31}$ \\
$2.5\times 10^{10}$   &  $0.3289\times 10^{-31}$   &  $0.1465\times 10^{-31}$   &  $0.4754\times 10^{-31}$ \\
$6.0\times 10^{10}$   &  $0.4427\times 10^{-31}$   &  $0.1995\times 10^{-31}$   &  $0.6422\times 10^{-31}$ \\
$1.0\times 10^{11}$   &  $0.5357\times 10^{-31}$   &  $0.2377\times 10^{-31}$   &  $0.7734\times 10^{-31}$ \\
$2.5\times 10^{11}$   &  $0.7320\times 10^{-31}$   &  $0.3247\times 10^{-31}$   &  $0.1057\times 10^{-30}$ \\
$6.0\times 10^{11}$   &  $0.9927\times 10^{-31}$   &  $0.4377\times 10^{-31}$   &  $0.1430\times 10^{-30}$ \\
$1.0\times 10^{12}$   &  $0.1179\times 10^{-30}$   &  $0.5196\times 10^{-31}$   &  $0.1699\times 10^{-30}$ \\
% $0.2500\times 10^{13}$   &  $0.1631\times 10^{-30}$   &  $0.7192\times 10^{-31}$   &  $0.2350\times 10^{-30}$ \\
% $0.6000\times 10^{13}$   &  $0.2161\times 10^{-30}$   &  $0.9568\times 10^{-31}$   &  $0.3118\times 10^{-30}$ \\
		%\hline
	\end{tabular}
	\label{tab:nuN4}
\end{center}
\end{table}

\begin{table}
\caption{Charged-current and neutral-current cross sections and their 
sum for $\bar{\nu} N$ interactions according to the 
CTEQ4--DIS distributions.}

\begin{center}
	\begin{tabular}{cccc}
		%\hline
		$E_{\nu}\hbox{ [GeV]}$ & $\sigma_{\mathrm{CC}}$ [cm$^2$] & 
		$\sigma_{\mathrm{NC}}$ [cm$^2$] & $\sigma_{\mathrm{tot}}$ [cm$^2$] \\
		\hline
	$1.0\times 10^{1}$  &  $0.3936\times 10^{-37}$  &  $0.1381\times 10^{-37}$  &  $0.5317\times 10^{-37}$ \\
	$2.5\times 10^{1}$  &  $0.9726\times 10^{-37}$  &  $0.3403\times 10^{-37}$  &  $0.1313\times 10^{-36}$ \\
	$6.0\times 10^{1}$  &  $0.2287\times 10^{-36}$  &  $0.7982\times 10^{-37}$  &  $0.3085\times 10^{-36}$ \\
	$1.0\times 10^{2}$  &  $0.3747\times 10^{-36}$  &  $0.1307\times 10^{-36}$  &  $0.5054\times 10^{-36}$ \\
	$2.5\times 10^{2}$  &  $0.9154\times 10^{-36}$  &  $0.3193\times 10^{-36}$  &  $0.1235\times 10^{-35}$ \\
	$6.0\times 10^{2}$  &  $0.2153\times 10^{-35}$  &  $0.7531\times 10^{-36}$  &  $0.2906\times 10^{-35}$ \\
	$1.0\times 10^{3}$  &  $0.3542\times 10^{-35}$  &  $0.1243\times 10^{-35}$  &  $0.4785\times 10^{-35}$ \\
	$2.5\times 10^{3}$  &  $0.8548\times 10^{-35}$  &  $0.3026\times 10^{-35}$  &  $0.1157\times 10^{-34}$ \\
	$6.0\times 10^{3}$  &  $0.1922\times 10^{-34}$  &  $0.6896\times 10^{-35}$  &  $0.2612\times 10^{-34}$ \\
	$1.0\times 10^{4}$  &  $0.3008\times 10^{-34}$  &  $0.1091\times 10^{-34}$  &  $0.4099\times 10^{-34}$ \\
	$2.5\times 10^{4}$  &  $0.6355\times 10^{-34}$  &  $0.2358\times 10^{-34}$  &  $0.8713\times 10^{-34}$ \\
	$6.0\times 10^{4}$  &  $0.1199\times 10^{-33}$  &  $0.4570\times 10^{-34}$  &  $0.1656\times 10^{-33}$ \\
	$1.0\times 10^{5}$  &  $0.1683\times 10^{-33}$  &  $0.6515\times 10^{-34}$  &  $0.2334\times 10^{-33}$ \\
	$2.5\times 10^{5}$  &  $0.2909\times 10^{-33}$  &  $0.1158\times 10^{-33}$  &  $0.4067\times 10^{-33}$ \\
	$6.0\times 10^{5}$  &  $0.4667\times 10^{-33}$  &  $0.1901\times 10^{-33}$  &  $0.6568\times 10^{-33}$ \\
	$1.0\times 10^{6}$  &  $0.6051\times 10^{-33}$  &  $0.2493\times 10^{-33}$  &  $0.8544\times 10^{-33}$ \\
	$2.5\times 10^{6}$  &  $0.9365\times 10^{-33}$  &  $0.3929\times 10^{-33}$  &  $0.1329\times 10^{-32}$ \\
	$6.0\times 10^{6}$  &  $0.1393\times 10^{-32}$  &  $0.5930\times 10^{-33}$  &  $0.1986\times 10^{-32}$ \\
	$1.0\times 10^{7}$  &  $0.1734\times 10^{-32}$  &  $0.7423\times 10^{-33}$  &  $0.2476\times 10^{-32}$ \\
	$2.5\times 10^{7}$  &  $0.2542\times 10^{-32}$  &  $0.1100\times 10^{-32}$  &  $0.3642\times 10^{-32}$ \\
	$6.0\times 10^{7}$  &  $0.3622\times 10^{-32}$  &  $0.1578\times 10^{-32}$  &  $0.5200\times 10^{-32}$ \\
	$1.0\times 10^{8}$  &  $0.4430\times 10^{-32}$  &  $0.1937\times 10^{-32}$  &  $0.6367\times 10^{-32}$ \\
	$2.5\times 10^{8}$  &  $0.6278\times 10^{-32}$  &  $0.2762\times 10^{-32}$  &  $0.9040\times 10^{-32}$ \\
	$6.0\times 10^{8}$  &  $0.8696\times 10^{-32}$  &  $0.3836\times 10^{-32}$  &  $0.1253\times 10^{-31}$ \\
	$1.0\times 10^{9}$   &  $0.1050\times 10^{-31}$  &  $0.4641\times 10^{-32}$  &  $0.1514\times 10^{-31}$ \\
	$2.5\times 10^{9}$   &  $0.1464\times 10^{-31}$  &  $0.6489\times 10^{-32}$  &  $0.2113\times 10^{-31}$ \\
	$6.0\times 10^{9}$   &  $0.2011\times 10^{-31}$  &  $0.8931\times 10^{-32}$  &  $0.2904\times 10^{-31}$ \\
	$1.0\times 10^{10}$  &  $0.2406\times 10^{-31}$  &  $0.1066\times 10^{-31}$  &  $0.3472\times 10^{-31}$ \\
	$2.5\times 10^{10}$  &  $0.3286\times 10^{-31}$  &  $0.1465\times 10^{-31}$  &  $0.4751\times 10^{-31}$ \\
	$6.0\times 10^{10}$  &  $0.4481\times 10^{-31}$  &  $0.1995\times 10^{-31}$  &  $0.6476\times 10^{-31}$ \\
	$1.0\times 10^{11}$  &  $0.5335\times 10^{-31}$  &  $0.2377\times 10^{-31}$  &  $0.7712\times 10^{-31}$ \\
	$2.5\times 10^{11}$  &  $0.7306\times 10^{-31}$  &  $0.3247\times 10^{-31}$  &  $0.1055\times 10^{-30}$ \\
	$6.0\times 10^{11}$  &  $0.9854\times 10^{-31}$  &  $0.4377\times 10^{-31}$  &  $0.1423\times 10^{-30}$ \\
	$1.0\times 10^{12}$  &  $0.1165\times 10^{-30}$  &  $0.5195\times 10^{-31}$  &  $0.1685\times 10^{-30}$ \\
%	$0.2500\times 10^{13}$  &  $0.1615\times 10^{-30}$  &  $0.7192\times 10^{-31}$  &  $0.2334\times 10^{-30}$ \\
%	$0.6000\times 10^{13}$  &  $0.2166\times 10^{-30}$  &  $0.9568\times 10^{-31}$  &  $0.3123\times 10^{-30}$ \\
		%\hline
	\end{tabular}
	\label{tab:anuN4}
\end{center}
\end{table}

\narrowtext
\begin{table}[tbp]
%	\centering
	\caption{Upward $\mu^{+}+\mu^{-}$ event rates per year arising from 
	$\nu_{\mu}N$ and $\bar{\nu}_{\mu}N$ interactions in rock, 
for a detector with effective area $A = 0.1\km^{2}$ and muon energy 
threshold $E_\mu^{\mathrm{min}}=1\tev$.  The rates are shown 
integrated over all angles below the horizon and restricted to ``nearly 
horizontal'' nadir angles $60^{\circ} < \theta < 90^\circ$.  }
	\begin{center}
\begin{tabular}{ldd} %\hline
  & \multicolumn{2}{c}{nadir angular acceptance} \\
Flux &  $0^\circ < \theta < 90^\circ$ &  $60^{\circ} < \theta < 90^\circ$  \\
\hline
 ATM \cite{volkova} &  1100. &  570.  \\
 ATM \cite{volkova} + charm \cite{prs} &  1100. &  570.  \\
 AGN-SS91 \cite{stecker} & 500. & 380. \\
 AGN-M95 ($p\gamma$) \cite{karlm} &  31.  &    18.    \\
 AGN-P96 ($p\gamma$) \cite{proth} &  45.  &   39.    \\ 
 %AGN-P96 ($p\gamma$ and $pp$) \cite{proth} &  2100. &    1200.  \\ 
 GRB-WB \cite{waxb} &   12.  &    8.1  \\ 
 TD-SLSC \cite{slsc}  &   0.005 &    0.0046 \\ 
 TD-WMB12 \cite{branden} & 0.50 & 0.39 \\
 TD-WMB16 \cite{branden} & 0.00050 & 0.00039 \\
\end{tabular}
\end{center}
	\label{tab:eth1}
\end{table}

\begin{table}[tbp]
	\centering
\caption{Upward $\mu^{+}+\mu^{-}$ event rates per year arising from 
$\nu_{\mu}N$ and $\bar{\nu}_{\mu}N$ interactions in rock, for a 
detector with effective area $A = 0.1\km^{2}$ and muon energy 
threshold $E_\mu^{\mathrm{min}}=10\tev$.  The rates are shown 
integrated over all angles below the horizon and restricted to ``nearly 
horizontal'' nadir angles $60^{\circ} < \theta < 90^\circ$.}
	\begin{center}
\begin{tabular}{ldd} %\hline
  &  \multicolumn{2}{c}{nadir angular acceptance} \\
Flux &   $0^\circ < \theta < 90^\circ$ &   $60^{\circ} < \theta < 90^\circ$ \\
\hline
 ATM \cite{volkova} &  17. &   10. \\
 ATM \cite{volkova} + charm \cite{prs} & 19. & 11. \\
 AGN-SS91 \cite{stecker} & 270. & 210. \\
 AGN-M95 ($p\gamma$) \cite{karlm} &  5.7  &    4.3  \\
 AGN-P96 ($p\gamma$) \cite{proth} &   28.  &  25.  \\ 
 %AGN-P96 ($p\gamma$ and $pp$) \cite{proth} &   440.  &  280.  \\ 
 GRB-WB \cite{waxb} &    5.4  & 4.0 \\ 
\end{tabular}
\end{center}
	\label{tab:eth10}
\end{table}

\begin{table}[tbp]
	\centering
\caption{Upward $\mu^{+}+\mu^{-}$ event rates per year arising from 
$\nu_{\mu}N$ and $\bar{\nu}_{\mu}N$ interactions in rock, for a 
detector with effective area $A = 0.1\km^{2}$ and muon energy 
threshold $E_\mu^{\mathrm{min}}=100\tev$.  The rates are shown 
integrated over all angles below the horizon and restricted to 
``nearly horizontal'' nadir angles $60^{\circ} < \theta < 90^\circ$.}
	\begin{center}
\begin{tabular}{ldd} %\hline
  &  \multicolumn{2}{c}{nadir angular acceptance} \\
Flux &   $0^\circ < \theta < 90^\circ$ &   $60^{\circ} < \theta < 90^\circ$ \\
\hline
 ATM \cite{volkova} &  0.13 &   0.09 \\
 ATM \cite{volkova} + charm \cite{prs} & 0.21 & 0.16 \\
 AGN-SS91 \cite{stecker} & 85. & 73. \\
 AGN-M95 ($p\gamma$) \cite{karlm} &  1.6  &    1.5  \\
 AGN-P96 ($p\gamma$) \cite{proth} &   13.  &  12.  \\ 
% AGN-P96 ($p\gamma$ and $pp$) \cite{proth} &   31.  &  24.  \\ 
 GRB-WB \cite{waxb} &    1.2  & 1.0 \\ 
\end{tabular}
\end{center}
	\label{tab:eth100}
\end{table}

\mediumtext
\begin{table}
\caption{Downward $\mu^{+}+\mu^{-}$ events per year arising from 
$\nu_{\mu}N$ and $\bar{\nu}_{\mu}N$ interactions in $1\km^{3}$ of 
water.}
\begin{center}
		\begin{tabular}{lddd}
		%\hline
		 & \multicolumn{3}{c}{Muon-energy threshold, 
		 $E_{\mu}^{\mathrm{min}}$} \\
		 Flux & $100\tev$ & $1\pev$ & $3\pev$ \\
		\hline
ATM \cite{volkova} & 0.85 & 0.0054 & 0.00047 \\
ATM \cite{volkova} + charm \cite{prs} & 2.6 & 0.050 & 0.0076 \\
AGN-SS91 \cite{stecker} & 520. & 120. & 42. \\
AGN-M95 ($p\gamma$) \cite{karlm} & 16. & 11. & 8.7  \\
AGN-P96 ($p\gamma$) \cite{proth} & 100. & 50. & 31. \\
GRB-WB  \cite{waxb} & 7.7 & 1.9 & 0.93 \\
TD-SLSC  \cite{slsc} & 0.037 & 0.032 & 0.029 \\
TD-WMB12 \cite{branden} & 1.1 & 0.74 & 0.58 \\
TD-WMB16 \cite{branden} & 0.00087 & 0.00050 & 0.00035 \\
		%\hline
		\end{tabular}
\end{center}
\label{tab:hith}
\end{table}

\widetext
\begin{table}[tbp]
%	\centering
\caption{Downward resonant $\bar{\nu}_{e}e\rightarrow W^{-}$ events 
per steradian per year for a detector with effective volume 
$V_{\mathrm{eff}}=1\km^{3}$.  Also shown are the downward (upward) 
potential background rates from $\nu_{\mu}N$ and $\bar{\nu}_{\mu}N$ 
interactions induced by neutrinos with $E_{\nu}>3\pev$.}
	\begin{center}
\begin{tabular}{ldddddd} %\hline
        %\hline
 Flux & $\bar{\nu}_{e}e \rightarrow \bar{\nu}_{\mu}\mu$ &
 $\bar{\nu}_{e}e \rightarrow \hbox{hadrons}$ &
 \multicolumn{2}{c}{$(\nu_{\mu}, \bar{\nu}_{\mu})N$ CC} &
 \multicolumn{2}{c}{$(\nu_{\mu}, \bar{\nu}_{\mu})N$ NC}  \\
 \hline
 AGN-SS91 \cite{stecker} & 6. & 41. & 29. & (5.2)  & 13. & (2.3)   \\
 AGN-M95 ($p\gamma$) \cite{karlm} & 0.1  & 0.6  & 2.3 & (0.21)  & 1.1 & (0.095)   \\
 AGN-P96 ($p\gamma$) \cite{proth} & 1.2 & 7.8 & 12. & (1.6)  & 5.2 & (0.69)   \\
 GRB-WB  \cite{waxb} & 0.06 & 0.4 & 0.43 & (0.065)  &  0.19 & (0.029) \\
 TD-SLSC \cite{slsc}  &  0.001 & 0.0074 &  0.0059 & (0.00031)  & 0.0028 
 & (0.00014) \\
 %\hline
                \end{tabular}
        \end{center}
	\label{tab:glashow}
\end{table}
\mediumtext
\begin{table}
\caption{
Annual neutral-current $(\nu_e,\, \bar{\nu}_e)N$ and
charged-current $(\nu_\mu,\, \bar{\nu}_\mu)N$ event rates for the
Pierre Auger Cosmic Ray Observatory.
}
\begin{tabular}{ldddd} 

Flux & \multicolumn{2}{c}{$E_{\mathrm{sh}}>10^8\gev$} & 
\multicolumn{2}{c}{$E_{\mathrm{sh}}>10^9\gev$}  \\
 &$ (\nu_e,\, \bar{\nu}_e)N$ NC  & $ (\nu_\mu,\, \bar{\nu}_\mu)N$ CC  
 &$ (\nu_e,\, \bar{\nu}_e)N$ NC  & $ (\nu_\mu,\, \bar{\nu}_\mu)N$ CC  \\
\hline
AGN-SS91 \cite{stecker}  & 0.0045 & 0.019 & 0.000006 & 0.000024 \\
AGN-M95 ($p\gamma$) \cite{karlm} & 0.65  & 2.7 & 0.26 & 1.1     \\
AGN-P96 ($p\gamma$)  \cite{proth} & 0.74 & 3.1 & 0.13 & 0.53     \\
GRB-WB \cite{waxb} & 0.038 & 0.16 & 0.020 & 0.085  \\
TD-SLSC \cite{slsc} & 0.013 & 0.052 & 0.010 & 0.042 \\
TD-WMB12 \cite{branden} & 0.15 & 0.59 & 0.11 & 0.44 \\
TD-WMB16 \cite{branden} & 0.000026 & 0.00011 & 0.000011 & 0.000046 \\
                \end{tabular}
\label{tab:PAnue}
\end{table}

\mediumtext

\begin{table}[tbp]
%	\centering
\caption{Annual event rates in the Pierre Auger Cosmic Ray Observatory 
for horizontal air showers induced by $(\nu_e,\bar{\nu}_e)N$ 
charged-current interactions.}
	\begin{center}
\begin{tabular}{lldd} 
%\hline
%\hline
Flux & Parton distributions &  $E_{\mathrm{sh}}>10^8\gev$ & 
$E_{\mathrm{sh}}>10^9\gev$ \\ 
\hline
  & CTEQ4--DIS     & 0.15   &  0.00026 \\
AGN-SS91 \cite{stecker} & CTEQ3--DLA & 0.13 & 0.00022 \\
  & D$\_^{\prime}$ & 0.23 & 0.00051 \\
\hline
  & CTEQ4--DIS     & 6.1   &  3.3\\
AGN-M95 ($p\gamma$) \cite{karlm}  & CTEQ3--DLA & 5.3    &  2.8 \\
      & D$\_^\prime$ & 12.  &  7.5 \\
\hline
  & CTEQ4--DIS     & 8.9    &  2.6 \\
AGN-P96 ($p\gamma$)  \cite{proth} & CTEQ3--DLA & 7.9    &  2.2 \\
      & D$\_^\prime$ & 16.  &  5.4 \\
\hline
 & CTEQ4--DIS     & 0.31    &  0.18 \\
GRB-WB \cite{waxb}      & CTEQ3--DLA & 0.27    &  0.16 \\
      & D$\_^\prime$ & 0.67  &  0.45 \\
\hline
      & CTEQ4--DIS     & 0.068    &  0.061 \\
TD-SLSC \cite{slsc}      & CTEQ3--DLA & 0.056    &  0.051 \\
      & D$\_^\prime$ & 0.18  &  0.17 \\
\hline
         & CTEQ4--DIS & 0.85 & 0.71 \\
TD-WMB12 \cite{branden} & CTEQ3--DLA & 0.72 & 0.60 \\
        & D$\_^\prime$ & 2.1 & 1.9 \\
\hline
        & CTEQ4--DIS & 0.00024 & 0.00014 \\
TD-WMB16 \cite{branden} & CTEQ3--DLA & 0.00021 & 0.00012 \\
        & D$\_^\prime$ & 0.00049 & 0.00032 \\
%\hline
\end{tabular}
\end{center}
	\label{tab:auger}
\end{table}

\narrowtext

%\clearpage
%\newpage
\section*{Figures}

\newpage

\begin{references}
\bibitem{mufldet} To estimate these fluxes, we use the incident flux 
at the Earth's surface given in Eqn.\ (29) of GQRS96 and the 
Lipari--Stanev range formula of Ref.\ \cite{ls}.
\end{references}
\end{document}